\documentclass[iop]{emulateapj}
\usepackage{epsf}
\usepackage{multirow}
\usepackage{url}
\bibpunct{(}{)}{;}{a}{}{,}
\usepackage{color}
\usepackage{listings}

\newcommand{\hmpc}{\,h^{-1}\,{\rm Mpc}}
\newcommand{\hmpcc}{\,h^{3}\,{\rm Mpc}^{-3}}
\newcommand{\vvir}{{\rm V_{vir}}}
\newcommand{\mvir}{{\rm M_{vir}}}
\newcommand{\vmax}{{\rm V_{max}}}
\newcommand{\vpeak}{{\rm V_{peak}}}

\begin{document}
\title{On the prospect of using the maximum circular velocity of halos to 
encapsulate assembly bias in the galaxy-halo connection}
\author{Idit Zehavi\altaffilmark{1,*}, Stephen E.\ Kerby\altaffilmark{1}, Sergio Contreras\altaffilmark{2,$\dagger$}, Esteban Jim\'enez\altaffilmark{3}, Nelson Padilla\altaffilmark{3,4}, \\ and Carlton M.\ Baugh\altaffilmark{5}}

\altaffiltext{*}{Email: idit.zehavi@case.edu}
\altaffiltext{$\dagger$}{Email: sergio.contreras@dipc.org} 
\altaffiltext{1}{Department of Physics, Case Western Reserve University,
Cleveland, OH 44106, USA}
\altaffiltext{2}{Donostia International Physics Center, 20018 Donostia, Basque
Country, Spain}
\altaffiltext{3}{Instit\'uto Astrof\'{\i}sica, Pontifica Universidad Cat\'olica
de Chile, Santiago, Chile}
\altaffiltext{4}{Centro de Astro-Ingenier\'{\i}a, Pontificia 
 Universidad Cat\'olica de Chile, Santiago, Chile}
\altaffiltext{5}{Institute for Computational Cosmology, Department of Physics, 
Durham University, South Road, Durham, DH1 3LE, UK}

\begin{abstract}  
We investigate a conceptual modification of the halo occupation distribution
approach, using the halos' present-day maximal circular velocity, $\vmax$, as 
an alternative to halo mass. In particular, using a semi-analytic galaxy 
formation model applied to the Millennium WMAP7 simulation, we explore the 
extent that switching to $\vmax$ as the primary halo property incorporates the 
effects of assembly bias into the formalism.  We consider fixed number density 
galaxy samples ranked by stellar mass and examine the variations in the halo 
occupation functions with either halo concentration or formation time. We 
find that using $\vmax$ results in a significant reduction in the occupancy 
variation of the central galaxies, particularly for concentration. The 
satellites occupancy variation on the other hand increases in all cases.  We 
find effectively no change in the halo clustering dependence on concentration, 
for fixed bins of $\vmax$  compared to fixed halo mass. Most crucially, we 
calculate the impact of assembly bias on galaxy clustering by comparing the 
amplitude of clustering to that of a shuffled galaxy sample, finding that the 
level of galaxy assembly bias remains largely unchanged. Our results suggest 
that while using $\vmax$ as a proxy for halo mass diminishes some of occupancy 
variations exhibited in the galaxy-halo relation, it is not able to encapsulate
the effects of assembly bias potentially present in galaxy clustering.  The use
of other more complex halo properties, such as $\vpeak$, the peak value of 
$\vmax$ over the assembly history, provides some improvement and warrants 
further investigation.
\end{abstract}

\keywords{cosmology: galaxies --- cosmology: theory --- galaxies: clustering --- galaxies: evolution --- galaxies: halos --- galaxies: statistics -- large-scale structure of universe}

\vspace{0.2cm}
\section{Introduction}
\label{Sec:Intro}

In the standard cosmological framework, galaxies form, evolve and reside
in dark matter halos. It is of fundamental importance to understand the
relation between galaxies and dark matter halos.  How do the galaxies 
populate the halos? How do the properties of galaxies depend on halo mass or 
other characteristics of the halos?  What role does the environment have in 
the halo occupation?  These questions lie at the core of our understanding 
of galaxy formation. They  are also crucial if we are to take full advantage 
of the next generation of galaxy surveys aimed at measuring galaxy clustering 
with unparalleled accuracy. The cosmological constraints from these data will 
no longer be dominated by statistical  errors but rather the accuracy of our 
theoretical models. Understanding how galaxies relate to the underlying dark 
matter is thus essential for optimally using galaxies as a cosmological probe.

The formation and evolution of the dark matter halos is dominated by gravity 
and can be predicted accurately using high-resolution cosmological numerical 
simulations and analytic models. The formation of the galaxies and their 
relation to the dark matter halos, however, is more complex and depends on 
the detailed physical processes leading to the varied observed galaxy 
properties. Studying in detail the galaxy-halo connection is therefore of 
paramount importance (see \citealt{Wechsler17} for a review).

A powerful approach to explore galaxy formation within dark matter halos is 
semi-analytic modeling \citep[SAM; e.g.,][]{Cole94,Cole00,Benson03,Baugh06,Croton06,Bower06,Somerville08}. 
In such models, halos identified from high resolution $N$-body simulations 
are ``populated'' with galaxies using analytical prescriptions for the 
evolution of baryons through cosmic time. 
These models have been successful in reproducing many measured properties 
including the galaxy luminosity and stellar mass functions  
\citep[see, e.g.,][]{Croton06,Bower06,Fontanot09,Guo11,Guo13,Henriques15,Lagos18,Baugh18}.

Alternatively, hydrodynamic simulations follow the physical processes which 
govern the behavior of baryons by solving the fluid equations, while also 
modeling some of the unresolved processes with sub-grid prescriptions 
\citep[see, e.g.,][]{Somerville15}. Cosmological hydrodynamical simulations 
have started to play a major role in the study of galaxy formation and 
evolution. Two recent efforts, the Illustris Project \citep{Vogelsberger14,TNG}
and the EAGLE simulation \citep{Schaye15}, have set the state-of-the-art 
in hydrodynamical calculations. These ambitious simulations are still however 
significantly smaller than large-scale structure dark-matter only simulations
and harder to fine-tune to the observations.

A useful approach to empirically connect galaxies with dark matter halos
is the Halo Occupation Distribution (HOD) framework
\citep[e.g.,][]{Jing98,Benson00,Peacock00,Seljak00,Scoccimarro01,Berlind02,Berlind03,Cooray02,Yang03,Kravtsov04,Zheng05}.
The HOD formalism  characterizes the relationship between galaxies and halos 
in terms of the probability distribution, ${\rm P(N|\mvir)}$, that a halo 
of virial mass $\mvir$ contains ${\rm N}$ galaxies of a given type, together 
with the spatial and velocity distributions of galaxies inside halos. The key
ingredient is the halo occupation function, ${\rm \langle N(\mvir) \rangle}$, 
representing the average number of galaxies as a function of halo mass. 
The advantage of this approach is that it does not rely on assumptions 
about the (poorly understood) physical processes that drive galaxy formation
and can be directly constrained from the observations.

When considering the halo occupation function it is often useful to separate
the contribution of central galaxies, namely the main galaxy at the center of 
the halo, and that of the additional satellite galaxies that populate the 
halo \citep{Kravtsov04,Zheng05,Jimenez19}. Standard applications assume a 
cosmology and a parametrized form for the halo occupation functions motivated 
by predictions of SAMs and hydrodynamic simulations \citep[e.g.,][]{Zheng05}. 
The HOD parameters are then constrained using galaxy clustering measurements 
from large surveys, the galaxies abundance and the predicted halo clustering. 
The approach has been demonstrated to be a powerful theoretical tool to
study the galaxy-halo connection, transforming clustering measurements 
into a physical relation between galaxies and dark matter halos. It has 
been successful in explaining the shape of the galaxy correlation function, 
its dependence on galaxy properties and environmental dependence 
\citep[e.g.,][]{Zehavi04,Zehavi05b,Zehavi11,Berlind05,Abbas06,Skibba06,Tinker08,Coupon12}.  It has also become an increasingly popular method to create 
realistic mock catalogs, by populating halos in large $N$-body simulations
\citep[e.g.,][]{Manera15,Zheng16,Smith17,AEMULUS}, important for planning and 
analysis of current and upcoming surveys.

A central assumption in the standard applications of the HOD framework is that 
the galaxy content of halos depends only on the host halo mass.
The origins of this assumption is in the Press-Schechter formalism 
\citep{Press74,Lacey93} and the uncorrelated nature of random walks describing 
halo assembly, resulting in a correlation of the halo environment with its 
mass but not with its assembly history \citep{Bond91,White99,Lemson99}.
This assumption has been challenged by explicitly demonstrating in large
$N$-body simulations that the clustering of halos varies with halo formation
time, concentration, substructure occupation and spin, at fixed halo mass
\citep{Sheth04,Gao05,Gao07,Wechsler06,Jing06,Wetzel06,Lazeyras17,Sato19}, 
an effect that has been generally referred to as ``halo assembly bias''.

To what extent is the galaxy distribution impacted by the assembly bias of 
their host halos is an actively debated topic. If galaxy properties closely 
correlate with halo formation history this would lead to a dependence of the 
galaxy content on large-scale environment and a corresponding change in the 
amplitude of galaxy clustering on large scales. The latter is commonly 
referred to as ``galaxy assembly bias'' (a misnomer referring to the 
manifestation of halo assembly bias in the galaxy distribution;  GAB 
hereafter). The predictions for GAB have been explored with simulated galaxies 
\citep[e.g.,][]{Zhu06,Croton07,Zu08,Zentner14,Chaves16,Romano17,Xu19}, 
while the observational evidence for it remains inconclusive and controversial.
Several suggestive detections have been put forward 
\citep{Cooper10,Wang13b,Lacerna14b,Watson15,Hearin15,Miyatake16,Montero17}  
while other studies indicate the impact of assembly bias is small 
\citep{Abbas06,Blanton07,Tinker08,Lacerna14a,Lin16,ZuMan16,Walsh19} and that 
previous claimed detections are due to systematics 
\citep[e.g.,][]{Campbell15,Zu16,Sin17,Tinker17a,Sunayama19}. 

If significant, such GAB can have direct implications for interpreting 
galaxy clustering using the HOD framework \citep{Pujol14,Zentner14}.
The existence of GAB essentially implies a dependence of the halo occupation 
functions on the secondary halo parameters in addition to mass. This
``occupancy variation'' (\citealt{Zehavi18}; Z18 hereafter) is essentially
the crucial link between halo assembly bias and GAB;  both halo assembly bias
and occupancy variation are required to produce GAB.   

Z18 explore the predicted occupancy variation in SAMs applied to the 
Millennium simulation, focusing on the dependence of the galaxy content 
of halos on large-scale environment and on halo formation time. They 
find distinct occupancy variations with central galaxies in denser regions
preferentially occupying lower-mass halos. A similar, but significantly 
stronger, trend is found with halo age, where early-formed halos are more 
likely to host central galaxies at lower halo mass. A reverse trend is seen 
for the occupation of satellites, with early-forming halos having fewer 
satellites.
Z18 investigate the stellar mass-halo mass relation for central galaxies
in the SAM, finding a clear dependence on secondary halo properties, e.g., 
at fixed halo mass, older halos tend to host more massive galaxies, which 
drive the centrals occupancy variation.
They also examine the resulting GAB, arising from the combined effect of 
halo assembly bias and the occupancy variations. Similar results for the 
occupancy variation have also been obtained  in the hydrodynamical simulations 
EAGLE, Illustris and IllustrisTNG \citep{Artale18,Bose19}.

\citet[hereafter C19]{Contreras19} extended the work of Z18 to study the
cosmic evolution of assembly bias and occupancy variation in the 
\citet{Guo13} SAM. They explore the redshift range between 0 and 3, considering 
galaxy samples selected by either stellar mass or star formation rate (SFR), 
and selecting halos by both halo formation time and halo concentration.  
At the present epoch, both halo concentration and halo formation time
produce similar features when examining either halo clustering or the occupancy
variation. These two halo properties however exhibit different evolutionary
scenarios. The GAB signature monotonically decreases when going to higher 
redshift (and for lower number density samples), reversing its sense in some 
instances.

Building on the work of Z18 and C19, we set out here to explore a conceptual 
modification of the HOD approach which shifts from halo mass to an alternative 
proxy, aimed at encapsulating assembly bias in the galaxy-halo connection.
We also draw upon abundance matching techniques (e.g.,
\citealt{Conroy06,Reddick13,Chaves16,Lehmann17}). Such methods typically 
associate dark matter (sub)halos with galaxies using a monotonic relation
between a galaxy property (such as stellar mass or luminosity) and a halo
property like its infall halo mass, its  maximum circular velocity, $\vmax$,
or the peak value of $\vmax$ across the assembly of the halo, $\vpeak$. 
These are expected to have a tighter relation with the galaxy properties
compared to the virial mass of the halo, $\mvir$.

Given the distinct trends with concentration (or halo age) in the galaxy-halo 
connection and their crucial role in producing the occupancy variations 
(Z18), we choose $\vmax$ as the alternate halo property. $\vmax$, the maximal
value of the circular velocity inferred from the halo mass distribution, 
effectively characterizes the gravitation potential.  Using the 
\citet{Guo13} SAM applied to the Millennium WMAP7 simulation, we explore 
the extent to which switching from virial mass to the halo's maximum circular 
velocity in the HOD is able to encapsulate the main features of assembly bias. 
We consider all three aspects of this phenomenon: occupancy variation, GAB, 
and halo assembly bias.  While some aspects are certainly improved, we find
(spoiler alert) that, regrettably, $\vmax$ is unable to ``capture'' the 
essence of assembly bias and improve upon the use of $\mvir$ in measuring GAB.
Still, we think it is educational and useful to go through this exercise.

The overall motivation and potential applications of this work are many. 
First, we can obtain further insight into assembly bias and on 
how best to capture its essential features. Beyond the increased theoretical
understanding of this complex phenomenon, had this conceptual change to
the HOD successfully encapsulated assembly bias, it would have provided
varied practical applications. The modified HOD could then have been 
utilized as an efficient tool for creating mock catalogs from simulations, 
which would reproduce realistic galaxy clustering while incorporating 
assembly bias. Moreover, it could have been potentially invaluable for 
inferring the galaxy-halo connection from galaxy clustering free of the 
effects of assembly bias.

The outline of the rest of the paper is as follows. Section 2 describes the
galaxy samples we use. In Section 3 we present the occupancy variation with
halo concentration and age when using $\vmax$ instead of halo mass. In Section
4 we examine the relation between stellar mass and the different halo 
properties. Section 5 presents our results for galaxy assembly bias, and
we conclude in Section 6. Appendix A investigates halo assembly bias with
$\vmax$. Appendix B includes additional results for SFR-selected samples
and samples at higher redshift. Finally, Appendix C shows some results
when alternatively using $\vpeak$. 

\vspace{0.2cm}
\section{The Galaxy Samples}
\label{Sec:Samples}

\vspace{0.1cm}
\subsection{Simulation and Galaxy Formation Model}

We use the Millennium WMAP7 N-body simulation \citep{Guo13}, which is similar
to the original Millennium simulation \citep{Springel05} but with cosmological
parameters consistent with the seven-year WMAP data \citep{Komatsu11}.  The 
simulation has a comoving boxsize of $500 \hmpc$ on a side and follows $2160^3$ 
particles with a mass resolution of $9.36 \times 10^8 h^{-1} {\rm M}_{\odot}$. 
Multiple outputs of the simulation are available at different snapshots from 
redshift $50$ to the present day. At each snapshot, halos are identified using 
a friends-of-friends algorithm \citep{Davis85}, and {\tt SUBFIND}  
\citep{Springel01} is then run on these to identify subhalos. Halo merger 
trees are constructed by following the evolution of the halos and subhalos 
with time. 

Semi-analytic modeling is a fundamental methodology to model the evolution 
of baryons over cosmic time in a cosmological context 
\citep[see, e.g.,][]{Baugh06, Lacey16}.  SAMs aim to model the main physical 
processes involved in galaxy formation and evolution, grafted onto a halo 
merger tree. Some of these processes include gas cooling, star formation, 
feedback from supernovae and active galactic nucleii, chemical enrichment, 
dark matter halo mergers and galaxy mergers. The SAM used in our work is that 
of \citet{Guo13}, which is a version of the {\tt L-GALAXIES} SAM code of the 
Munich group, based on \citet{Guo11} and applied to the Millennium WMAP7 
simulation. It is publicly available from the Millennium Archive{\let\thefootnote\relax\footnote{$^6$\,\,\url{http://gavo.mpa-garching.mpg.de/Millennium/}}}. 

\begin{figure}
\includegraphics[width=0.46\textwidth]{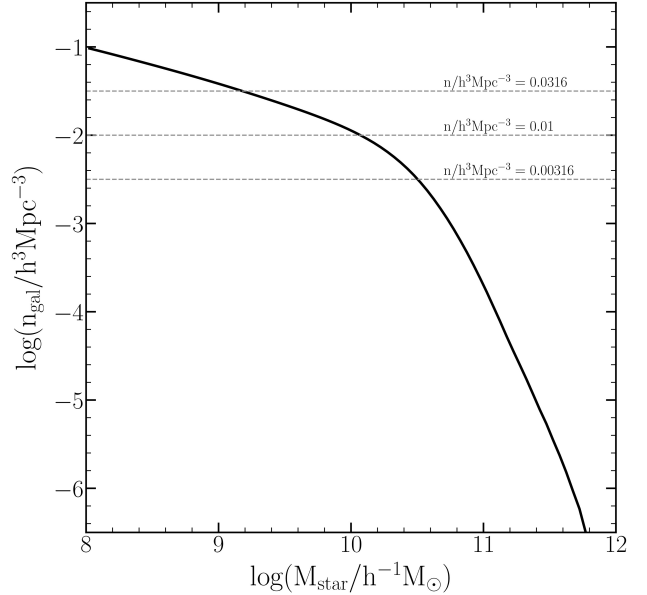}
\caption{The cumulative stellar mass function predicted by the \citet{Guo13}
SAM.  The dashed horizontal lines indicate the number densities of the samples
used in this work.}
\label{fig:CMF}
\end{figure}

\begin{figure*}
\includegraphics[width=0.48\textwidth]{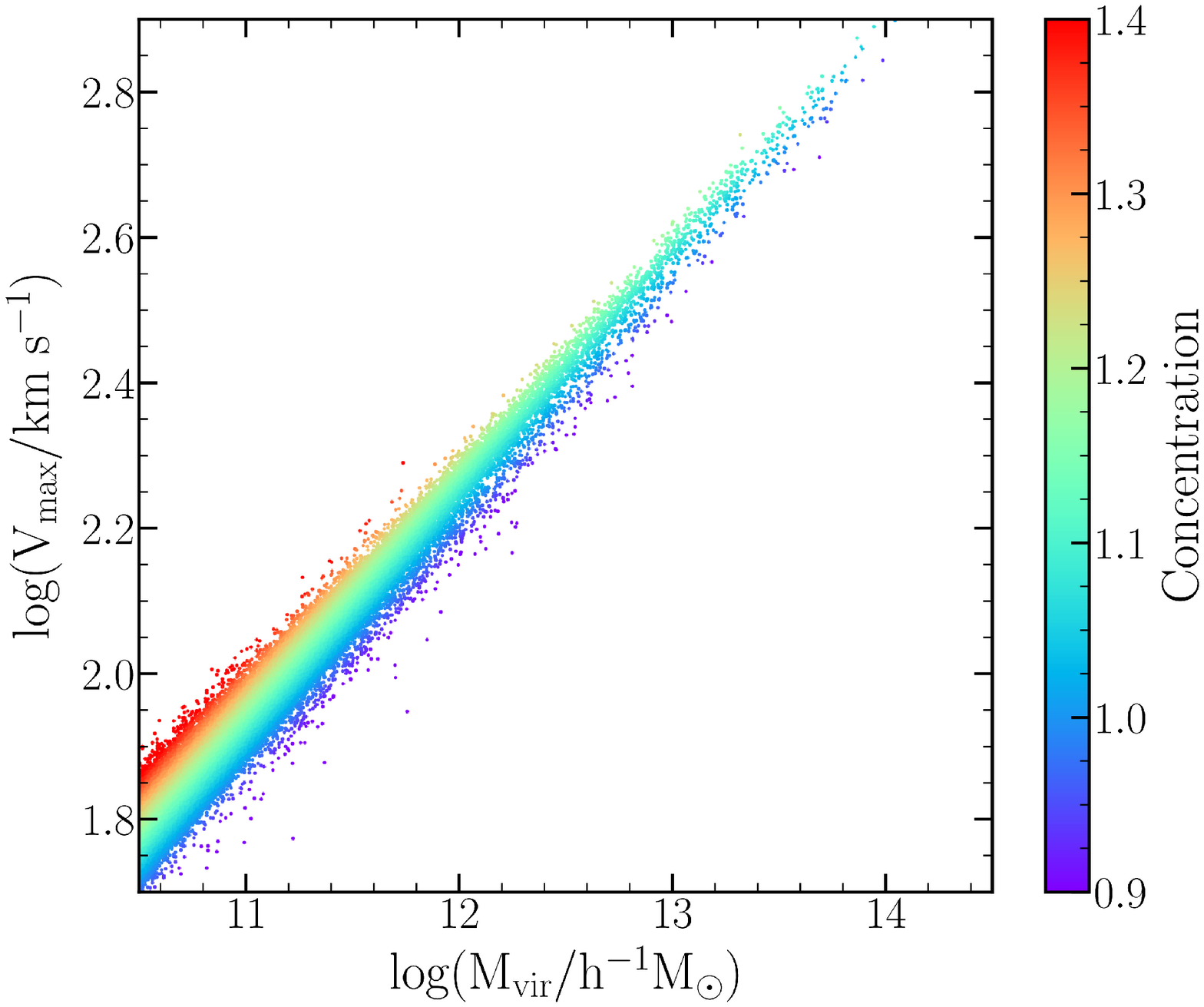}
\hspace{0.4cm}
\includegraphics[width=0.48\textwidth]{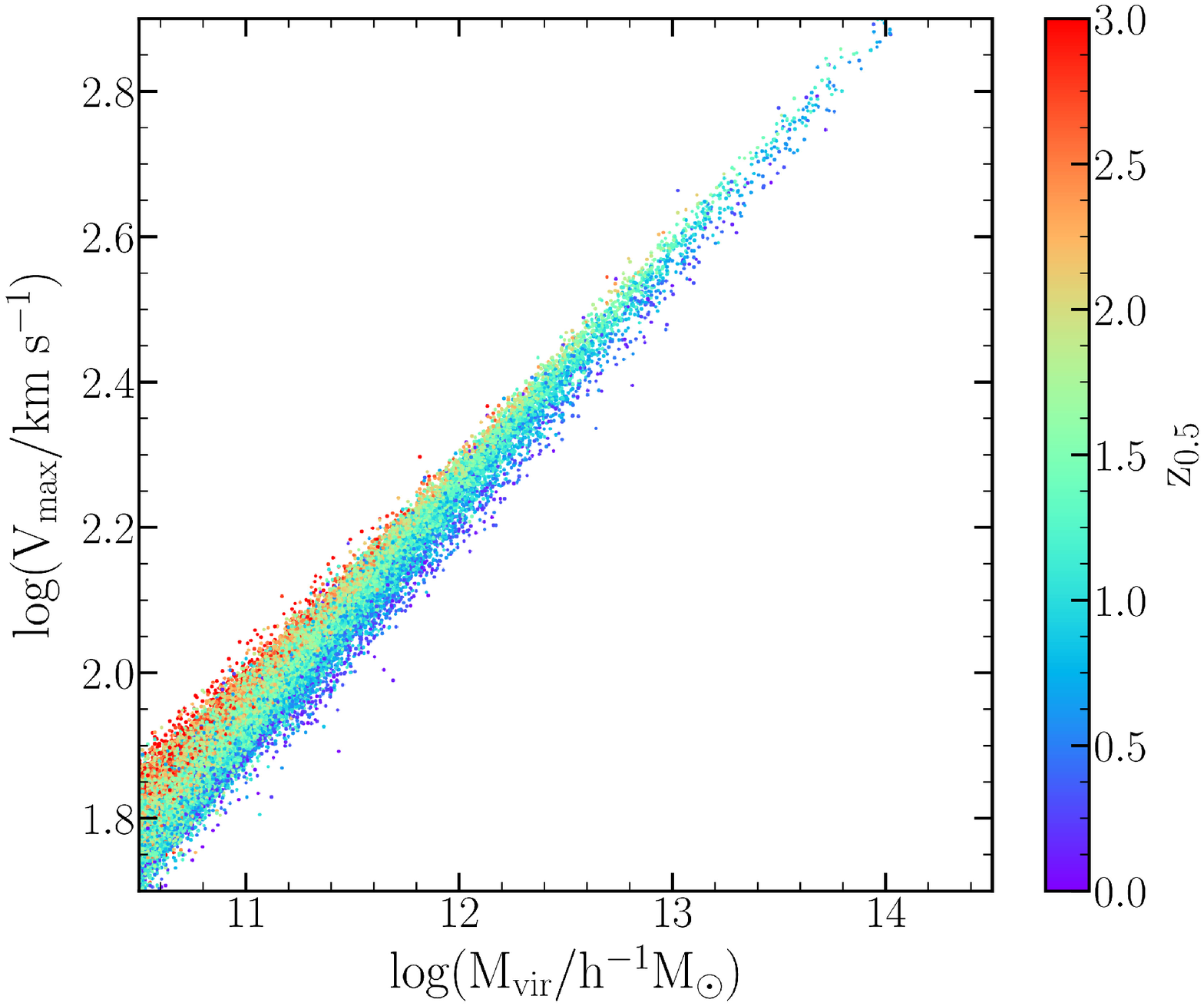}
\caption{The relation between $\vmax$ and $\mvir$, color coded by halo 
concentration (left) and by halo formation time (right). For clarity and to
avoid overcrowding, we plot only $1\%$ of the halos randomly chosen and 
ordered.}
\label{fig:VmaxMvir}
\end{figure*}

For the main part of our analysis we use galaxy samples with different
number densities, ranked by the stellar mass of the galaxies. We use three
number densities, $0.0316,\, 0.01,$ and  $0.00316 \hmpcc$, corresponding to
stellar mass thresholds of $\sim 1.6 \times 10^9,\, 1.2 \times 10^{10}$ and  
$3.4 \times 10^{10} h^{-1} {\rm M}_{\odot}$ , respectively.
The samples are approximately evenly spaced in logarithmic number density, 
and follow the choices made in Z18. Figure~\ref{fig:CMF} shows the cumulative
stellar mass function and the three number density cuts used. The galaxies
selected in each case are those to the right of the intersection with the 
corresponding horizontal line. Pertinent results for galaxy samples selected 
by ranked SFR are shown in Appendix~B.

\vspace{0.1cm}
\subsection{Halo Properties}

We consider as the mass of the halo the virial mass, namely the mass
enclosed within the virial radius of the halo corresponding to a density of
200 times the critical density of the universe. This choice, which we
denote here as $\mvir$,  is largely a matter of convention but has been shown 
to roughly correspond to the boundary at which halos are in approximate 
dynamical equilibrium \citep[e.g.,][]{Cole96}.  We hereafter interchangeably
refer to $\mvir$ as the halo mass (but cf. \citealt{Jiang14} for an alternate
mass definition). The circular velocity profiles of the halos obtained from
the dark matter distribution, ${\rm V_c^2 = GM(r)/r}$,  have a well defined
maximum, which serves as a natural halo size scale denoted as $\vmax$.
This quantity is robustly defined for simulated halos and effectively 
characterizes the depth of the gravitational potential 
\citep[e.g.,][]{Bullock01,Conroy06,Diemand11}. We consider here in detail 
the use of $\vmax$ as a proxy for halo mass.

For the secondary parameters, we use the halo concentration and halo formation
time,  following C19. These are regarded as two fundamental
parameters related to the halo assembly often used in assembly bias studies
(e.g., \citealt{Gao05,Wechsler06,Gao07,Mao17}; C19; \citealt{Bose19}). 
The halo concentration characterizes the density profile. It is canonically 
defined as ${\rm C_{vir} = r_{vir}/r_{s}}$, where ${\rm r_{vir}}$ is the 
virial radius of the halo and $r_{\rm s}$ is the inner transitional radius 
appearing in the \citet{NFW} profile, at which the density profile changes 
slope. It is often alternatively defined as the ratio between $\vmax$ and 
$\vvir$, where $\vvir$ is the virial velocity of the halo, $\vvir \equiv 
{\rm V_c(r_{vir})}$. We use the latter definition here, which is directly 
calculable from simulation data and does not require any model fitting,
and can be applied to the lowest-mass halos considered, which might have
too few particles to fit a density profile \citep{Bullock01,Gao07,Diemand11}. 

The formation time of the halo is defined as the redshift at which the main 
progenitor of the halo first reaches half of its current mass, denoted as
${\rm z}_{0.5}$.  We calculate it from the halo merger trees of the simulation, 
linearly interpolating the halo mass among the time snapshots available. This 
definition has been very commonly used in assembly bias studies (e.g., 
\citealt{Gao05,Gao07,Croton07}; Z18; \citealt{Han19}). As we use the halo 
formation time simply to rank the halos as early-forming or late-forming we 
do not anticipate any dependence on the specific definition (cf., the study of 
\citealt{Li08}).

Figure~\ref{fig:VmaxMvir} shows the relation between $\vmax$ and $\mvir$ for the
halos in the simulation, color coded either by halo concentration or formation
time.  We plot only 1\% of the halos, randomly chosen and ordered, to avoid 
overcrowding. As expected, there is a tight relation between the two, with the 
scatter arising distinctly from the differences in concentration. For fixed 
halo mass, $\vmax$ is directly related to the concentration (by definition), 
with more concentrated halos corresponding to larger $\vmax$.  A similar 
relation is noted when color coding the halos by formation time, albeit with 
some scatter in the dependence on halo age, which arises from the scatter 
between concentration and halo age.

\vspace{0.2cm}
\section{Occupancy Variations}
\label{Sec:OV}

We now proceed to examine the halo occupation functions and their variations 
with halo concentration and age. For each of these properties, we rank the 
halos by the secondary property in narrow (0.1 dex) bins of halo mass and 
identify the 20\% extremes of the distribution. This factors out the halo mass 
dependence on these parameters and allows us to examine the occupancy variations
for halos of the same mass.  We also do the same for fine (0.04 dex) bins in 
$\vmax$. We have verified for both cases that our results are insensitive to 
the exact binning choice.  Given the tight relation between $\mvir$ and
$\vmax$ there is significant overlap between the identification binned by
either mass proxy.  

\begin{figure*}
\includegraphics[width=0.48\textwidth]{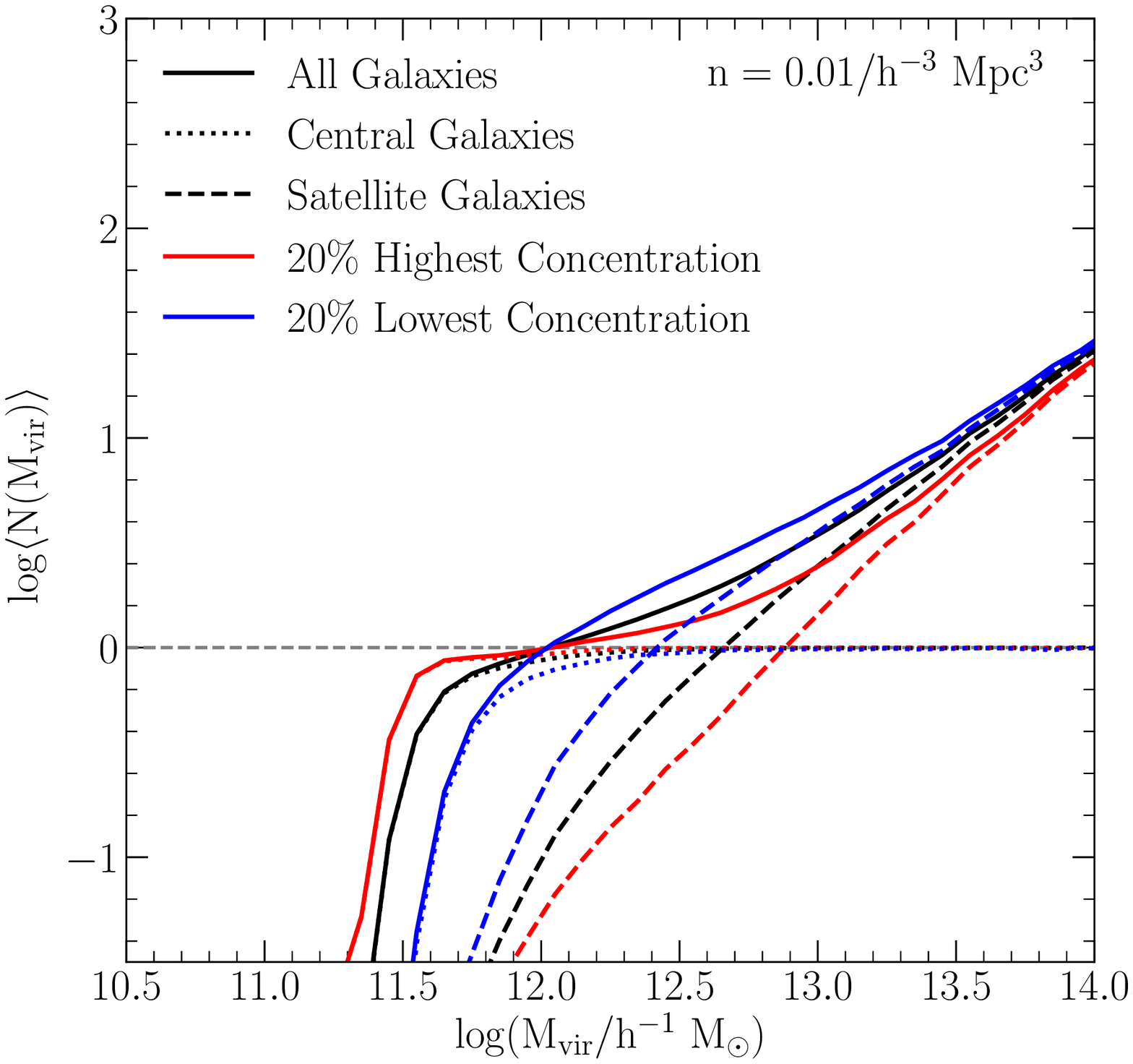}
\hspace{0.4cm}
\includegraphics[width=0.48\textwidth]{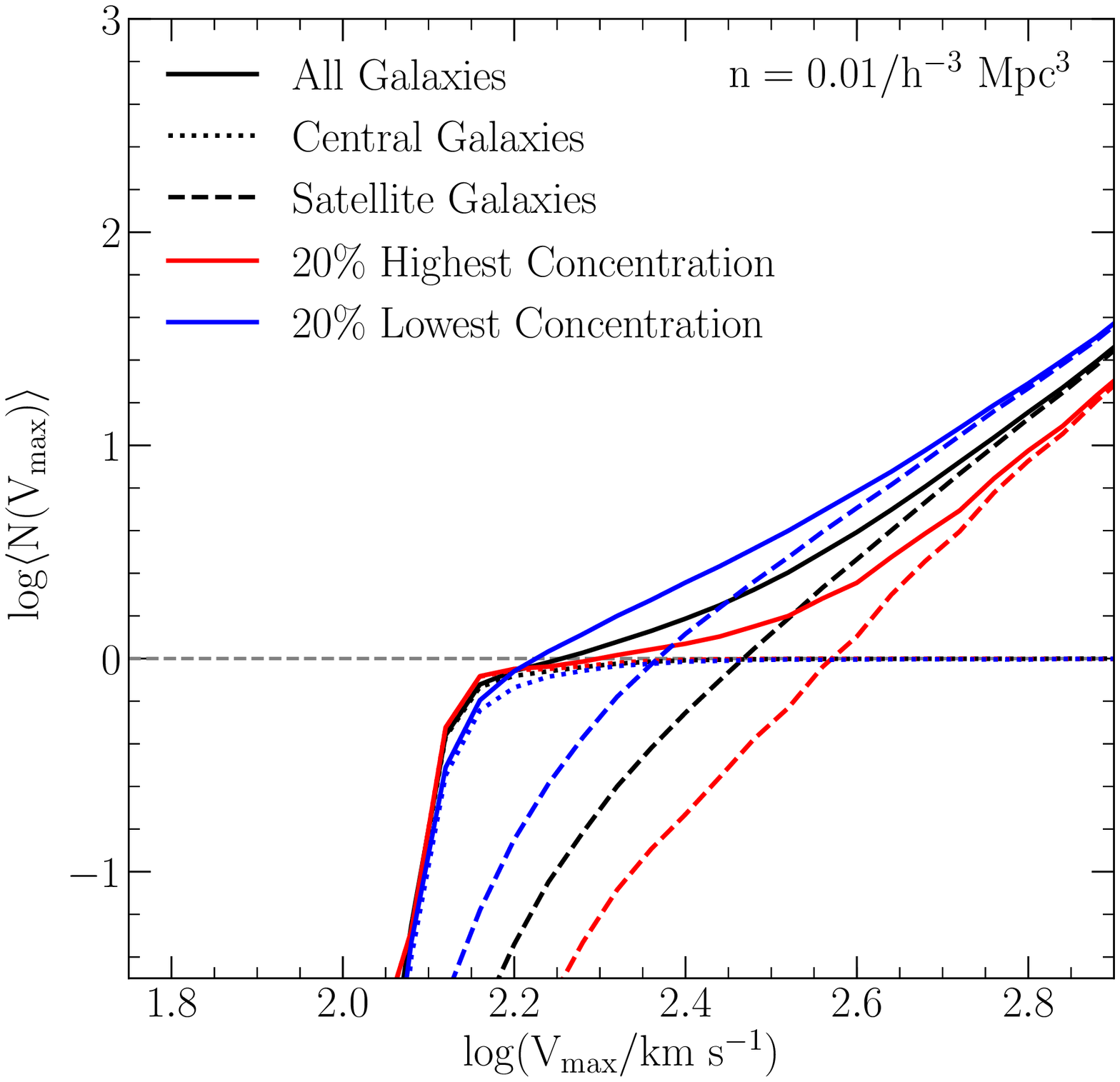}
\caption{Halo occupation functions for the $n=0.01 \hmpcc$ galaxy sample, for 
the full galaxy sample (black), for the galaxies in the 20\% most concentrated
halos (red) and 20\% least concentrated halos (blue).  The occupation functions
are shown as a function of $\mvir$ on the left and $\vmax$ on the right.}
\label{fig:HOD}
\end{figure*}

Figure~\ref{fig:HOD} presents the halo occupation functions for the
$n=0.01 \hmpcc$ galaxy sample and their variation with concentration. The 
left-hand side shows the ``standard'' occupation functions as a function of
halo mass. We see that more concentrated halos start hosting central galaxies
at lower halo mass, while they host on average fewer satellites per halo.
These are in full agreement with previous results explored in detail by
Z18 and C19. The uncertainties on the occupation functions, estimated from 
jackknife resampling, are negligible over the range of halo masses plotted 
here, and hence they are not included. (They only start becoming noticeable 
for masses larger than $\sim 10^{14} h^{-1} {\rm M}_\odot$ where there are 
very few halos; See, e.g., Fig.~3 of Z18 or Fig.~5 of C19).

The right-hand panel of Fig.~\ref{fig:HOD} shows our new results for the
halo occupations now using $\vmax$ as our proxy for halo mass. We see 
that the general shape of the HOD remains the same. However, there are 
significant changes to the occupancy variation of central galaxies and 
satellites.  We find that the $\vmax$ occupancy variation for central 
galaxies is very nearly diminished in this case, with all halos, most 
concentrated ones and least concentrated halos exhibiting similar occupation 
by galaxies. The $\vmax$ occupation functions for satellites, however, 
exhibit larger differences (namely, increased occupancy variation).

\begin{figure}
\includegraphics[width=0.48\textwidth]{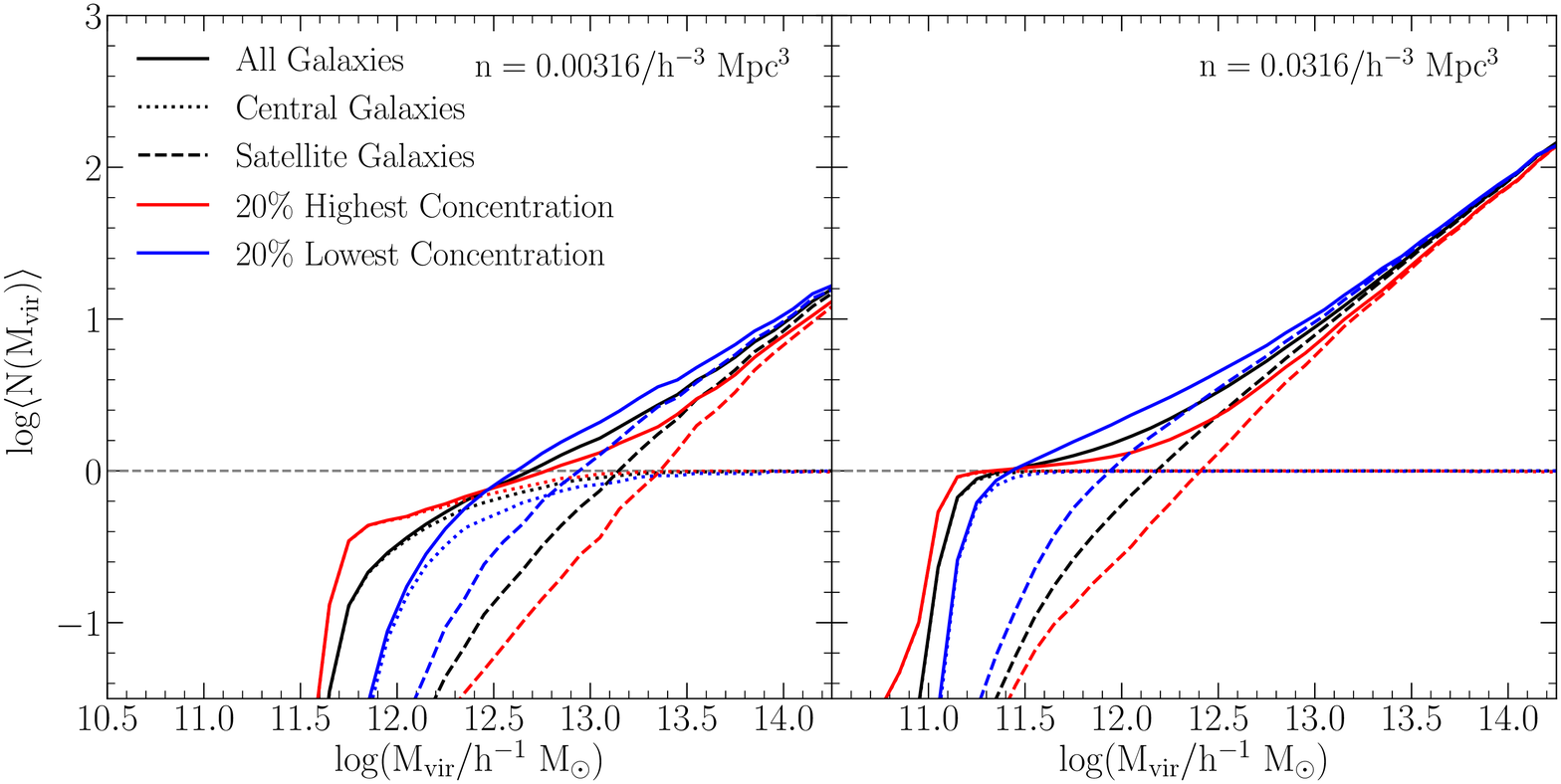}
\includegraphics[width=0.48\textwidth]{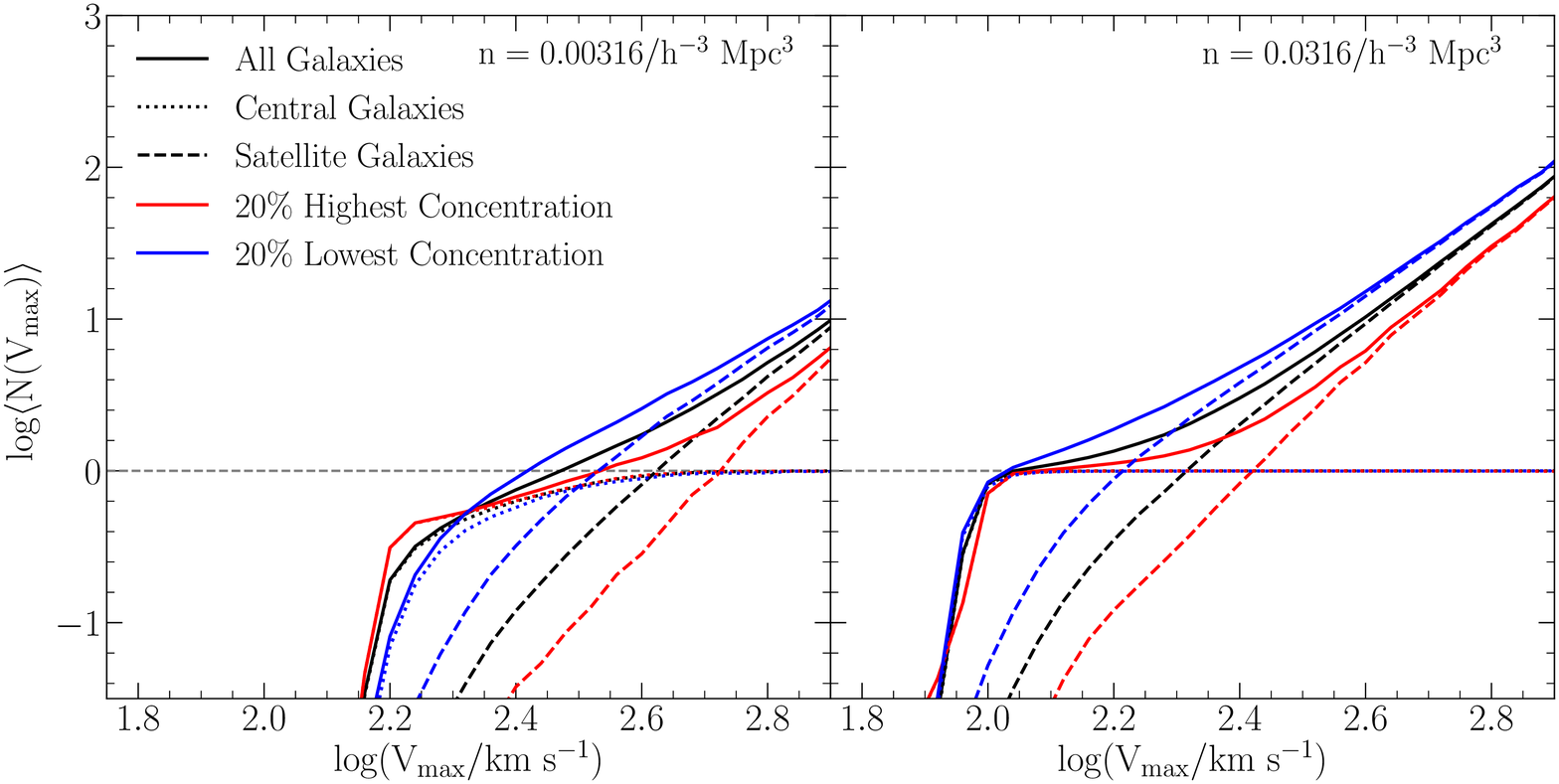}
\caption{The same as in Fig.~\ref{fig:HOD}, but for different number densities
samples, $n=0.00316 \hmpcc$ (left) and $n=0.0316 \hmpcc$ (right). The occupation
functions with $\mvir$ are shown on top and the ones with $\vmax$ on the bottom.}
\label{fig:HODdiffn}
\end{figure}

We also examine the differences in the occupancy variation for the other
galaxy samples in Figure~\ref{fig:HODdiffn}, finding some dependence
on the number density, or rather stellar mass threshold, of the sample. For 
the lower number density (more massive galaxies; left-hand side) switching 
from $\mvir$ to $\vmax$ only partially removes the central occupancy variation,
while it increases further the satellites occupancy variation. While for the 
higher number density (less massive galaxies; right-hand side) the central
occupancy variation is nearly overly compensated by the switch to $\vmax$, 
resulting in the least concentrated halos being occupied by central galaxies
at slightly lower $\vmax$ than the most concentrated halos. 

This change in behavior is not really unexpected and simply stems from the
dependence of the (standard) occupancy variation with number density, as 
exhibited in the top panels of Figure~\ref{fig:HODdiffn} and studied in detail 
in Z18. The halo mass occupancy variation for both centrals and satellite 
galaxies increases with stellar mass (i.e., with decreased number density). 
Switching from $\mvir$ to $\vmax$ amounts to a roughly constant shift in the 
mapping from $\mvir$ to $\vmax$, for the 20\% least/most concentrated halos, 
as can be seen in Fig.~\ref{fig:VmaxMvir}.  This results in the slight 
over-compensation, nearly full compensation and partial compensation of the 
$\vmax$ occupancy variation, respectively, for the three cases with 
decreasing number density. Thus the nearly diminished $\vmax$ occupancy 
variation seen in Fig.~\ref{fig:HOD} is somewhat coincidental.

\begin{figure*}
\includegraphics[width=0.96\textwidth]{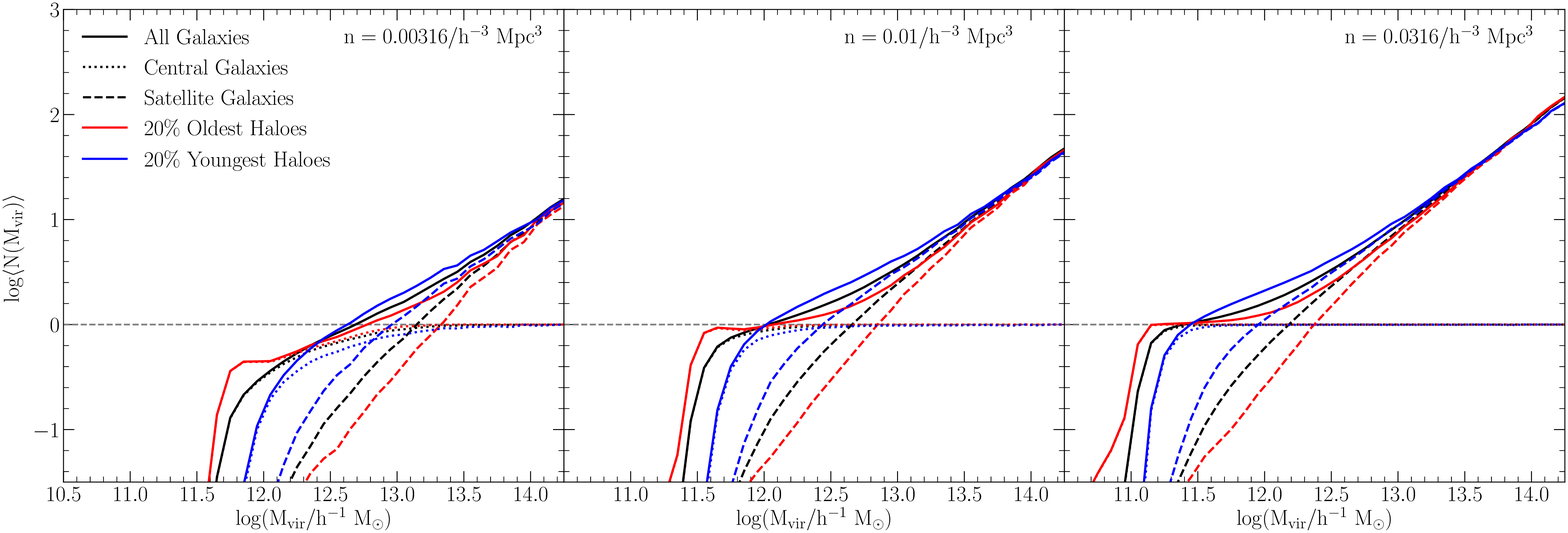}
\includegraphics[width=0.96\textwidth]{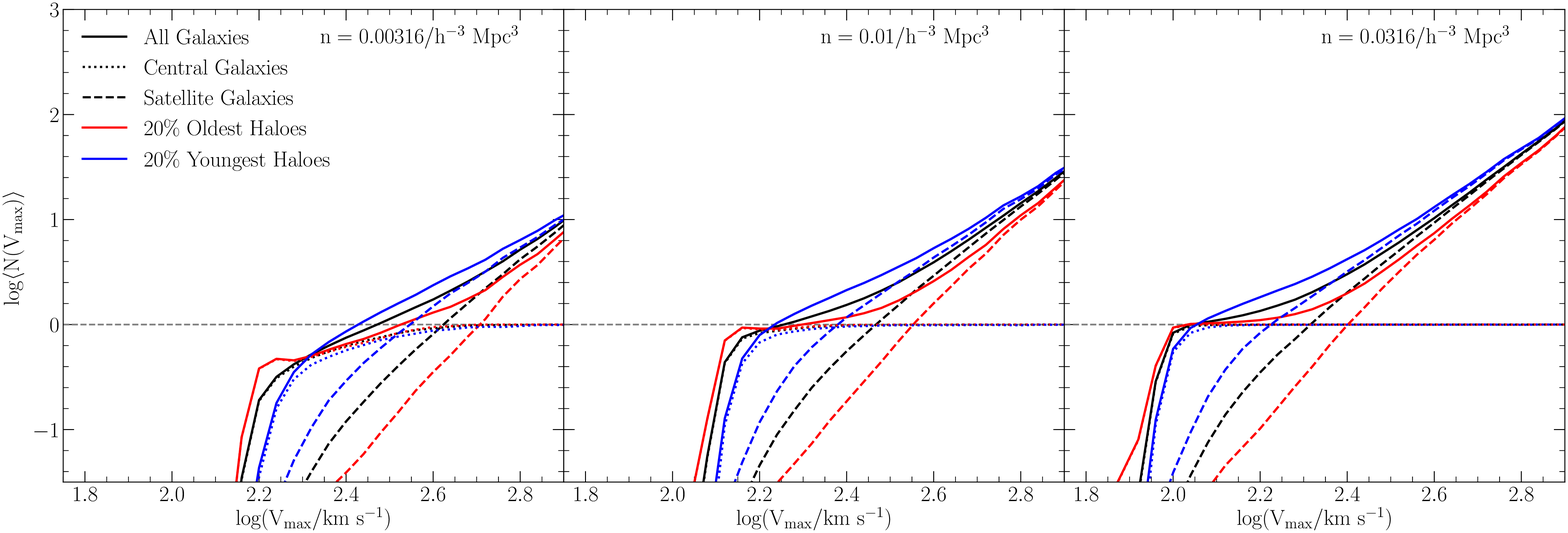}
\caption{The same as Fig.~\ref{fig:HOD} and \ref{fig:HODdiffn} but now for 
the occupancy variations with age, for all three number densities. The galaxies
in the 20\% oldest halos are shown in red, while the ones in the 20\% youngest
halos are shown in blue. The occupation functions with $\mvir$ are shown on 
top and the ones with $\vmax$ on the bottom.}
\label{fig:OVage}
\end{figure*}

To get a better sense of the amount of occupancy variations captured by
switching to $\vmax$,  we explore the variations also with regard to another
fundamental halo parameter, the halo formation time.  Figure~\ref{fig:OVage}
shows the occupation functions as a function of $\mvir$ and $\vmax$, for the
three number densities explored, now focusing on the 20\% early-formed halos 
and 20\% late-forming halos. When examining the standard HOD as a function
of $\mvir$, we find the well-studied variations with age (Z18; C19), with 
older halos more likely to host central galaxies at lower mass and to have 
fewer satellites. These variations are reduced when switching to $\vmax$ 
(bottom panels of Fig.~\ref{fig:OVage}), for all number densities,  but only 
partially so. This is in accord with the emerging understanding that galaxy 
assembly bias is not fully governed by a single parameter (or simple 
combination thereof; \citealt{Croton07,Xu19,Bose19}). The reduction noted here 
is likely due to the general correlation between halo formation time and 
concentration, and hence $\vmax$. This correlation of halo age with $\vmax$ 
is also seen in the right panel of Fig.~\ref{fig:VmaxMvir}. 

\begin{figure*}
\includegraphics[width=0.48\textwidth]{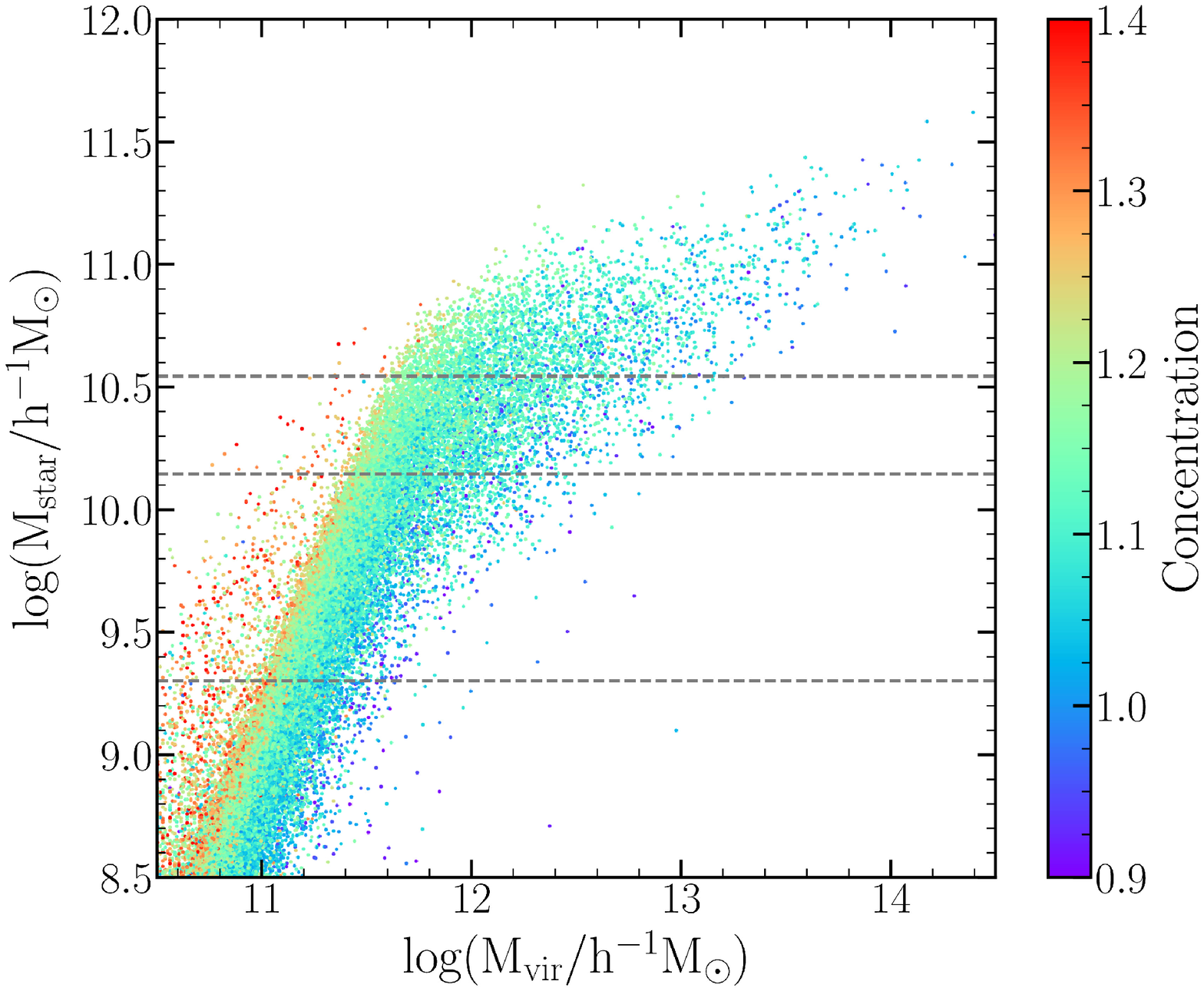}
\hspace{0.4cm}
\includegraphics[width=0.48\textwidth]{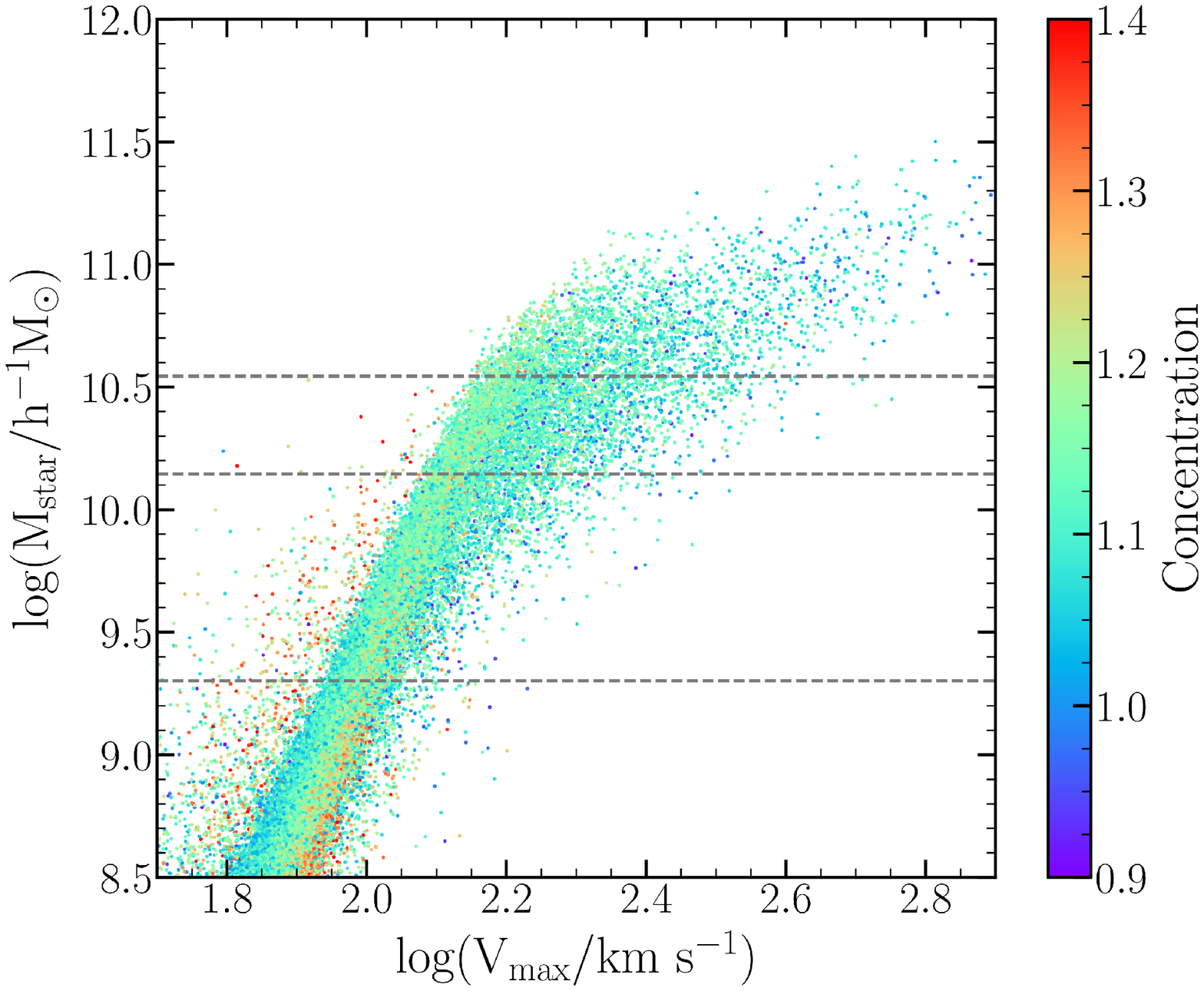}
\caption{\label{fig:SMHM}
The stellar mass of central galaxies as a function of host halo mass
(left) and $\vmax$ (right) for galaxies in the SAM applied to the Millennium 
WMAP7 simulation. Galaxies are color coded by their host halo concentration.
For clarity, we use a representative (randomly chosen and ordered) $1\%$ of
the central galaxies. The dashed horizontal lines shows the stellar mass 
thresholds used to define the galaxy samples.}
\end{figure*}

\vspace{0.2cm}
\section{Relation between Galaxy and Halo Properties}
\label{Sec:SMHM}

The study of Z18 first showed that the essential features of the centrals
occupancy variation stem from the dependence of the galaxies' stellar 
mass on secondary halo properties at fixed halo mass. They examined the 
stellar mass-halo mass relation (see their Fig.~9), finding that at fixed 
halo mass, more massive centrals tend to reside in older or denser halos.
This directly produces the changes in occupation we observe, and the level
of scatter in these secondary trends controls the strength of the occupancy
variation. Similar diagnostics were also utilized in other works such as
\citet{Matthee17,Tojeiro17,Artale18,Xu19}.

We revisit this relation in the context of our work. The left panel of
Figure~\ref{fig:SMHM} examines the relation between stellar mass and host halo 
mass for the central galaxies in the SAM, and the galaxies are color coded by 
the halo concentration. In an analogous fashion, we find here that at fixed 
halo mass, the more concentrated halos tend to host more massive central 
galaxies. The three horizontal lines mark the stellar mass thresholds used to 
define our samples. By examining the population of centrals above each 
threshold, we can visually see how the occupation variations come about, 
including the galaxies in the most-concentrated halos at lower halo mass and 
vice versa. The extent of the spread in the horizontal direction also directly 
correlates with the magnitude of the variations, producing larger variations 
for the higher stellar mass thresholds (lower number densities), as seen in 
Fig.~\ref{fig:HOD} and \ref{fig:HODdiffn}. We note, however, that the 
color-coding by concentration here is done globally, while when defining the 
samples for the occupation functions, the 20\% extremes are defined as a 
function of halo mass, so one should be careful when making the comparison.

The right panel of Fig.~\ref{fig:SMHM} shows the relation between  the
stellar mass of the central galaxies and $\vmax$ of their host halos, once
again color-coded by concentration. The stellar mass-$\vmax$ relation has
overall slightly less scatter, as expected (see \citealt{Chaves16,Matthee17}). 
In this case, we see a less-obvious trend with concentration. 
Examining the three cases separately (the three horizontal lines) we can
intuitively understand the result of each. For the middle sample, there
is a very mild trend with concentration. The top sample (lowest number
density) has in fact a larger horizontal scatter and trend with concentration
(after accounting for the different concentration markings noted above),
causing the remaining occupancy variation shown in Fig.~\ref{fig:HODdiffn}.
Going to the lowest stellar-mass threshold we in fact note that the
trend reverses with the lower-concentration halos appearing at lower 
halo mass than the higher concentration ones. Our highest number density
sample is just at the cusp of this change.

\vspace{0.2cm}
\section{Galaxy Assembly Bias}
\label{Sec:GAB}

Having investigated the behavior of the halo occupation functions and the
resulting occupancy variations with $\vmax$, we now proceed to see its impact
on galaxy clustering, namely GAB.  For completeness, we also provide a brief 
investigation of halo assembly bias in Appendix~A, showing the 
concentration-dependent halo clustering with $\mvir$ and $\vmax$.

To investigate the impact of assembly bias on galaxy clustering we compare
the clustering of galaxies in our sample to that of shuffled galaxy samples.
The shuffling follows the methodology of \citet{Croton07} (see also Z18 and 
C19), randomly reassigning the galaxy content of halos among halos of the 
same mass. More specifically, the central galaxies are randomly shuffled 
among halos within the same mass bin. The satellite galaxies are moved 
together with their original central galaxy, preserving the same distribution, 
and thus maintaining the same contribution to the correlation function from 
intra-halo pairs.  In order to investigate galaxy assembly bias with respect 
to $\vmax$, we perform an analogous shuffling procedure in $\vmax$ bins.
We present the results using 0.05 dex bins in $\mvir$ and 0.044 dex bins
in $\vmax$, but we checked different binnings and verified
that this makes no difference to our results.
 
The shufflings remove any dependence of the galaxy population on secondary
properties of the halos (or their assembly history) other than that inherent
in halo mass or $\vmax$.  The difference between the clustering of the 
original galaxy sample and the shuffled sample, then reflects the impact of
assembly bias on galaxy clustering, namely the level of GAB. In the idealized 
extreme case, if $\vmax$ were able to encapsulate all the occupancy variations 
exhibited by galaxies, there would be no difference between the clustering
of the original sample and the $\vmax$-shuffled one.

\begin{figure}
\includegraphics[width=0.48\textwidth]{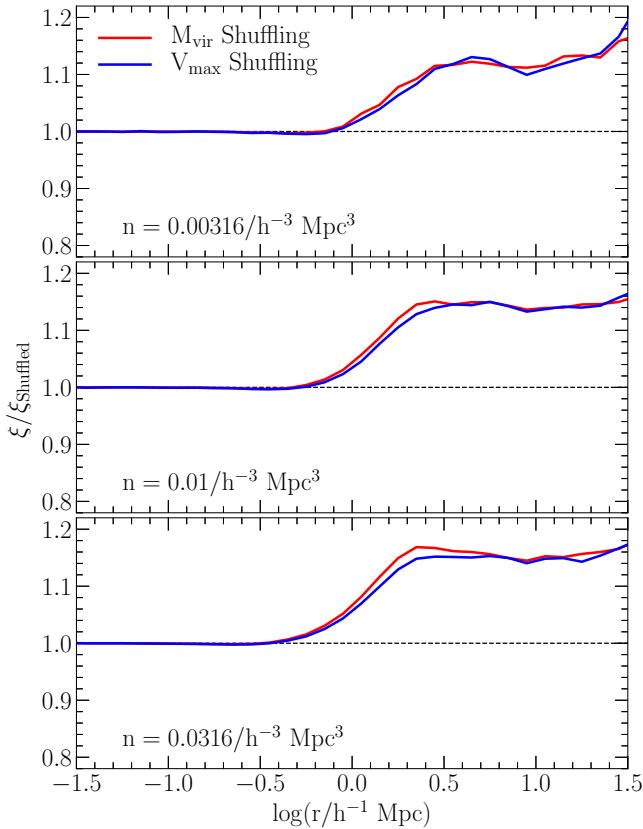}
\caption{The ratios between the correlation functions of the SAM galaxies and 
those of the corresponding shuffled galaxy samples (see text), showing the 
impact of galaxy assembly bias. These are shown for $\mvir$ (red) and $\vmax$ 
(blue) for three stellar-mass selected samples as labelled.}
\label{fig:GAB}
\end{figure}

Figure~\ref{fig:GAB} shows the resulting GAB signatures when shuffling 
either by $\mvir$ or $\vmax$, for the three stellar-mass selected samples  
studied.  The uncertainties on this clustering ratio, estimated from jackknife 
resampling, are again negligible over most of the range (and start becoming
noticeable only above separations of $20 \hmpc$;  see, e.g., Fig.~10 of Z18
or Fig.~10 of C19). The results for $\mvir$ are identical to the ones examined 
in C19, showing a $\sim$12\%-16\%  excess clustering due to assembly bias. This
arises from the combined effect of the occupancy variation and halo assembly
bias: central galaxies preferentially occupy the more concentrated halos
which are more strongly clustered.  The surprising, and rather disappointing,
results lies with the $\vmax$ case.  We find that in general the GAB associated
with $\vmax$ is comparable to that of $\mvir$.  This is somewhat puzzling 
given the significant reduction of the centrals occupancy variation when using 
$\vmax$.

\begin{figure}
\includegraphics[width=0.48\textwidth]{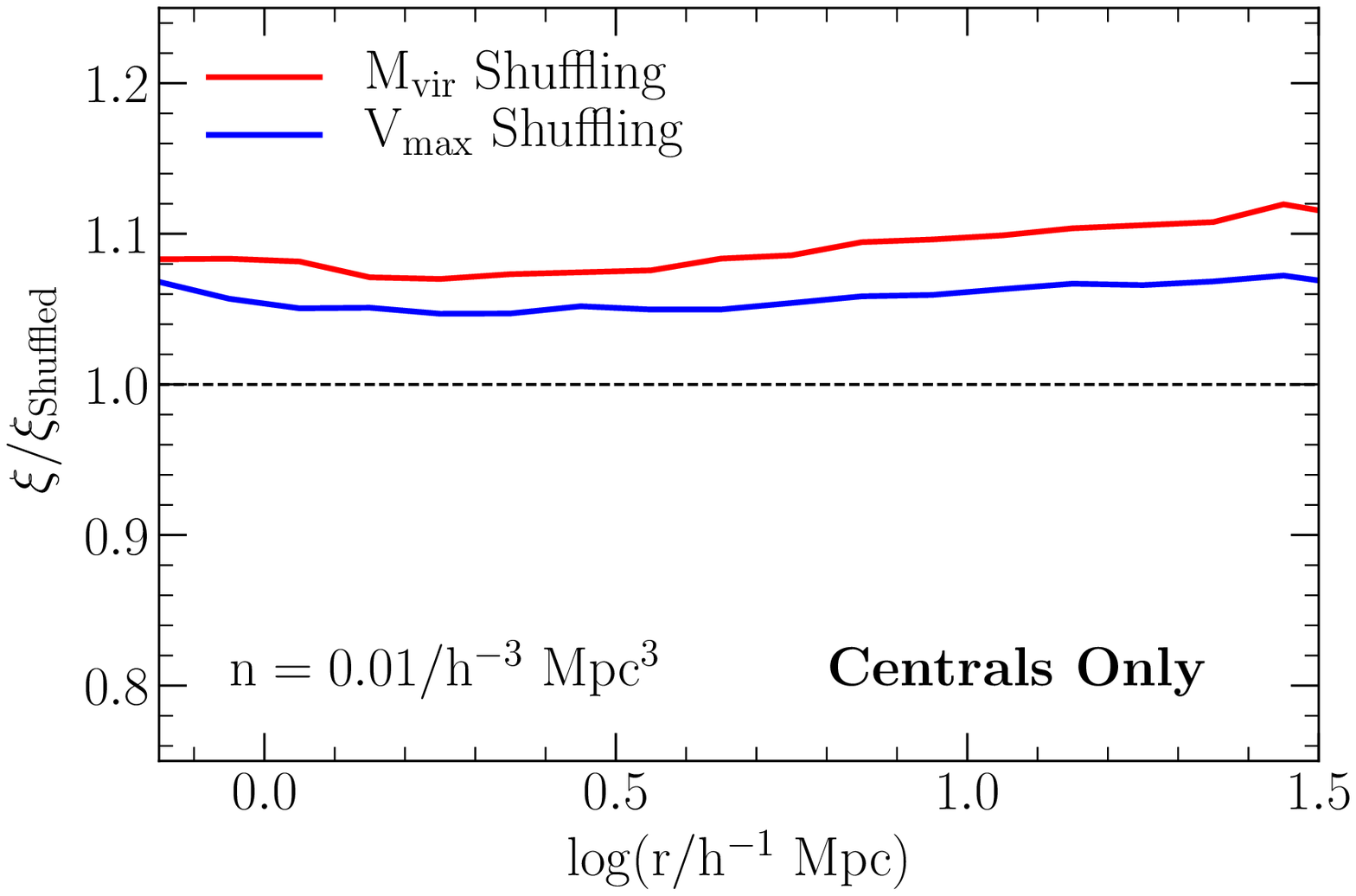}
\includegraphics[width=0.48\textwidth]{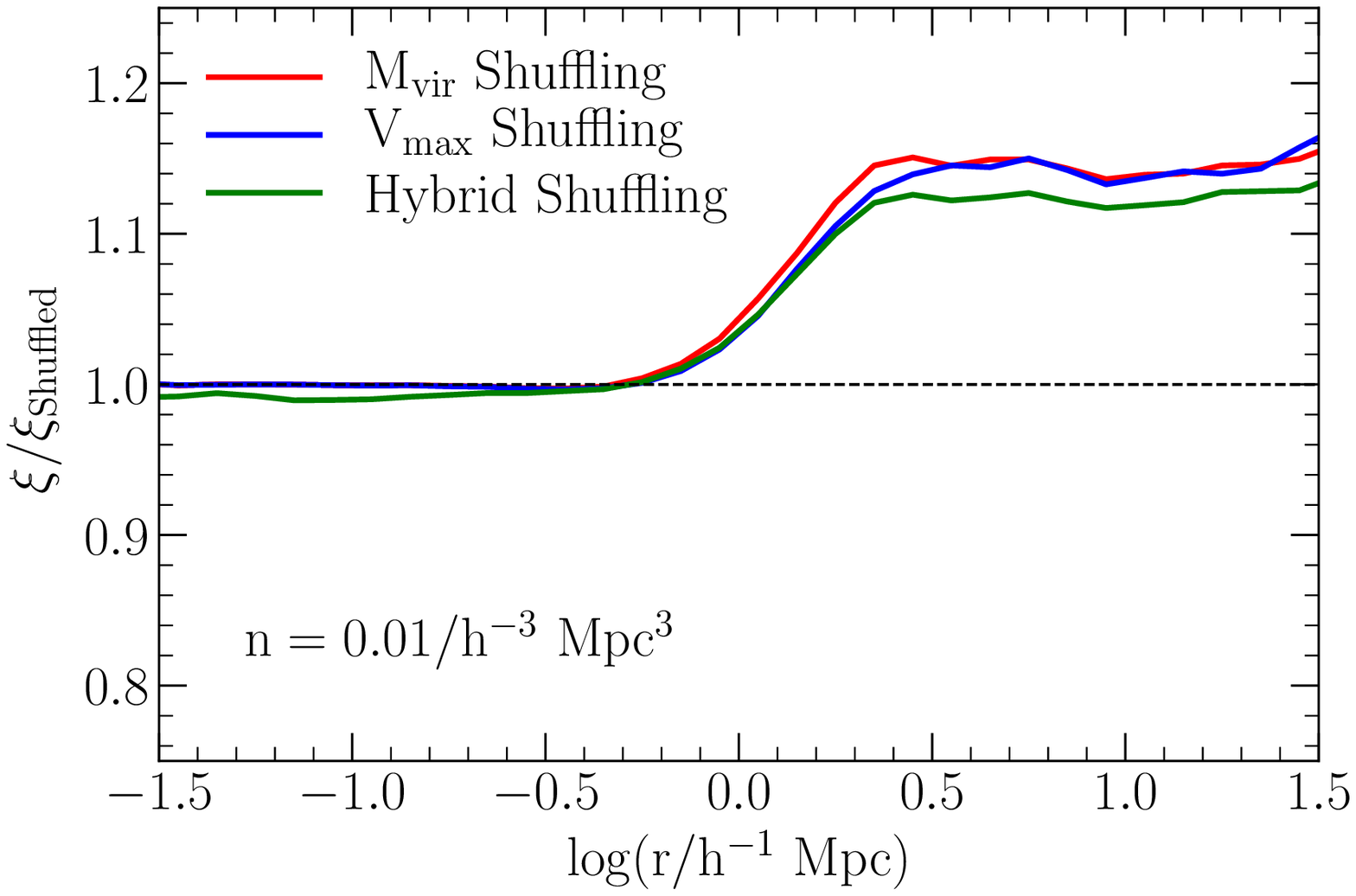}
\caption{Galaxy assembly bias measurements for only the central galaxies in the 
$n=0.01 \hmpcc$ sample (top)  and a hybrid shuffling method, where the 
centrals are shuffled by $\vmax$ and the satellites by $\mvir$ (bottom).}
\label{fig:GABtests}
\end{figure}

To further assess the situation, we repeat this analysis but now considering
only the central galaxies, the results of which are presented in the top panel
of Figure~\ref{fig:GABtests}.  Note that we only show here larger scales in 
the 2-halo regime, since without satellites there are no intra-halo pairs. We
see that when considering just the central galaxies there is a significant 
reduction in the GAB signal, of about $40\%$ at $10 \hmpc$ scale.
This reduction originates from the amount of assembly bias ``captured'' by
$\vmax$,  reflected in the much reduced occupancy variation with concentration
(shown in Fig.~\ref{fig:HOD}).  The occupancy variations however are
not fully captured by the switch to $\vmax$ when considering other parameters,
as demonstrated for halo age in Fig.~\ref{fig:OVage} (see also 
\citealt{Croton07,Matthee17}) and thus most of the GAB effect remains.

The lack of overall reduction in GAB for $\vmax$ compared to the standard
GAB signature for $\mvir$ exhibited in Fig.~\ref{fig:GAB} is likely due
to the significantly increased occupancy variation for satellites, which 
also contribute to the signal via central-satellite pairs 
\citep{Zu08}. It appears that the decreased centrals occupancy variation 
for $\vmax$ together with the increased satellites occupancy variation, and 
coupled with the roughly unchanged halo assembly bias (Appendix~A), result 
in a comparable level of GAB to the original measurement.

To remedy the increased contribution from the satellites occupancy variation
we attempt a hybrid shuffling scheme, where the central galaxies are shuffled 
by $\vmax$ while the satellites are shuffled by $\mvir$. The result of this 
hybrid shuffling is shown as the green line in the bottom panel of 
Fig.~\ref{fig:GABtests}, exhibiting a roughly $15\%$ reduction in GAB on large 
scales. The slight deviation of the hybrid case from a ratio of unity on small 
scales arises from changes to the contribution from central-satellite pairs 
due to the more complex shuffling procedure. One can envision improving this 
further by finding a different halo property that would diminish the satellites
occupancy variations, perhaps by utilizing the number of substructures. 
Alternatively, one may be able to develop a composite new parameter that will 
simultaneously improve both the centrals occupancy variation and the satellites 
one. However, attempts to do this at least in the context of halo assembly bias 
\citep[e.g.,][]{Mao17,Villarreal17,Xu18,Han19} have shown that this is largely 
unattainable, and such modeling approaches might be too complex to be practical 
in any case.

Another potential candidate is $\vpeak$, the peak value of $\vmax$ 
throughout the accretion history of the halo.  This parameter is often used
in the context of abundance matching methods 
\citep[e.g.,][]{Conroy06,Reddick13,Chaves16,Guo16} to better connect galaxies 
to halo substructure.  Its main advantage in such methodologies, however, is 
in connecting satellites to subhalos, which differs from the HOD approach
relating both central and satellite galaxies to the main host halo. Still,
recent claims have suggested that $\vpeak$ provides the tightest relation to
the stellar mass of galaxies \citep{He19} and is free from secondary 
dependences \citep{Xu19}, touting its use as a better proxy for halo mass.  
Remaining scatter in the relation is attributed to stochastic baryonic 
effects \citep{Matthee17,Kulier19}.

\begin{figure}
\includegraphics[width=0.48\textwidth]{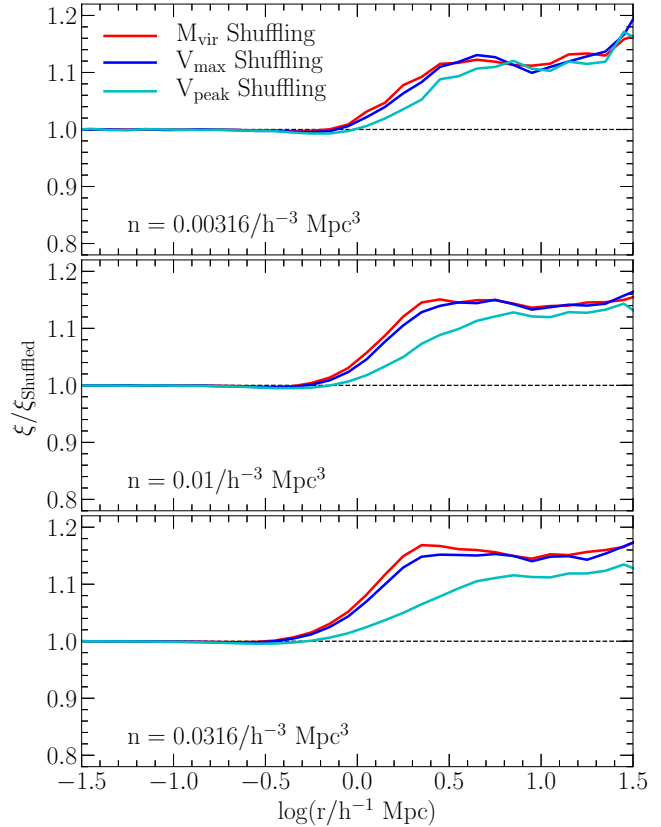}
\caption{Galaxy assembly bias measurements for $\vpeak$ (cyan) compared to the
results for $\mvir$ (red) and $\vmax$ (blue).}
\label{fig:Vpeak}
\end{figure}

Figure~\ref{fig:Vpeak} shows the resulting GAB, when using $\vpeak$
as the primary halo property instead of $\mvir$ or $\vmax$, for the three
different galaxy samples. Namely, we calculate the clustering of the galaxies
relative to a galaxy sample shuffled according to $\vpeak$, where $\vpeak$ is 
obtained from the merger tree of each halo. We see that using $\vpeak$ 
partially suppresses the GAB signature, mostly in the 1-halo to 2-halo 
transition region. This effect is more significant with increased number 
density, as more small galaxies are included.  Appendix~C shows further 
analysis of the relation of $\vpeak$ to other properties, suggesting 
that the differences arise primarily from low-mass, high-concentration and
older halos where $\vpeak$ varies from $\vmax$. This may be related to the
existence of splashback galaxies, i.e., galaxies whose halos had their mass 
accretion histories truncated due to a close encounter with a larger halo.
Such halos are thought to have a significant role in halo assembly bias of low 
mass halos \citep{Mansfield19}, and are likely to impact the intermediate
``transition'' scales, as these centrals likely reside just outside the virial
radius of the larger halo. We defer further investigation of this phenomenon
with higher-resolution simulations to future work.

We note, however, that even if $\vpeak$ can be utilized as a parameter 
which partially encapsulates assembly bias, determining it in simulations 
requires calculating the full merger trees of the halos with sufficient mass 
resolution.  In such a scenario, one will likely also have access to other 
sophisticated modeling techniques which may be more useful.  
We focused in this work on a property like $\vmax$ that would be readily 
available in most N-body simulations commonly used for creating large mock 
catalogs and for constraining cosmology. 

\vspace{0.2cm}
\section{Conclusion}
\label{Sub:Conc}

We use a state-of-the-art semi-analytic galaxy formation model applied to 
the Millennium simulation to study the prospects of a conceptual modification 
of the HOD approach, replacing the virial mass of the halo by its maximal 
circular velocity of the halo, $\vmax$.  The motivation is that this revised 
halo occupation function may encapsulate the effects of assembly bias into the
formalism, enabling more accurate modeling of the galaxy-halo connection
and galaxy clustering, and allow us to produce realistic mock catalogs that 
incorporate this effect.
We thus explore the different aspects of assembly bias, namely halo clustering
dependence on secondary parameters (halo assembly bias; see Appendix~A), the 
variation in the galaxy content of halos with these parameters (occupancy 
variation; \S~\ref{Sec:OV}) and the their impact on galaxy clustering relative
to a shuffled galaxy sample (galaxy assembly bias; \S~\ref{Sec:GAB}).
We mostly use here galaxy samples with fixed number density ranked by stellar 
mass at the present epoch, and for the secondary halo parameters we 
investigate the variation with halo concentration and halo formation time. To 
get a broader understanding of the origins of the occupancy variation, we 
also examine the relation between stellar mass and the different halo 
properties (\S~\ref{Sec:SMHM}).  Finally, we also investigate the potential 
of utilizing $\vpeak$, the peak value of $\vmax$ across the halo's assembly 
history as the proxy for halo mass.
  
The main conclusions from our work are summarized as follows:
\begin{itemize}

\item  Halo assembly bias, i.e. the dependence of halo clustering on
concentration, is largely unchanged when examined in bins of fixed $\vmax$
versus fixed halo mass. The same holds for $\vpeak$.

\item  Employing $\vmax$ significantly reduces the occupancy variation with 
halo concentration for central galaxies, however, it increases the satellites
occupancy variation.

\item  The centrals occupancy variation is partially reduced when using 
halo formation time as the secondary halo property. 

\item The change to $\vmax$ does not reduce the level of GAB, despite the 
reduction in the central occupancy variation.  The GAB signature remains 
essentially the same when using $\vmax$ or $\mvir$, irrespective of sample 
number density.

\item $\vmax$ does not prove to be a useful quantity also when examining
samples selected by SFR or when looking at $z=1$, with varying results.

\item Using $\vpeak$ slightly reduces the GAB signal, impacting mostly the
1-halo to 2-halo transition regime, an effect likely related to splashback 
galaxies. However, the large-scale effect remains largely unmitigated.
\end{itemize}

Assembly bias remains a challenge for contemporary models of galaxy 
clustering and the galaxy-halo relation. Perhaps a more intricate parameter
can better encapsulate the effects of galaxy assembly bias, however given
the complex nature of the phenomenon this might be hard to achieve and  
not-trivial to implement in practice. While the level of assembly bias in 
the Universe remains an open and debated issue, the results shown here can 
help inform theoretical modeling of it and attempts to determine it in 
observations. 

\smallskip
\acknowledgments
This work was made possible by the efforts of Gerard Lemson and colleagues at the German Astronomical Virtual Observatory in setting up the Millennium Simulation database in Garching. We acknowledge useful discussions with participants of the Shanghai Assembly Bias workshop in the final stages of this work.
IZ, SC, EJ \& NP acknowledge the hospitality of the ICC at Durham University.
IZ and SEK acknowledge support by NSF grant AST-1612085.
SC is supported by a Juan de la Cierva Formacion Fellowship (FJCI-2017-33816).
EJ acknowledges support from ``Centro de Astronom\'{i}a y Tecnolog\'{i}as Afines'' BASAL 170002. 
NP acknowledges support from Fondecyt Regular 1191813.
This project received financial support from the European Union's Horizon 
2020 Research and Innovation programme under the Marie Sklodowska-Curie grant 
agreement number 734374.

\vspace{0.2cm}
\appendix

\vspace{0.1cm}
\section{A. Halo Assembly Bias}
\label{Sec:HAB}

\begin{figure*}[h]
\includegraphics[width=0.48\textwidth]{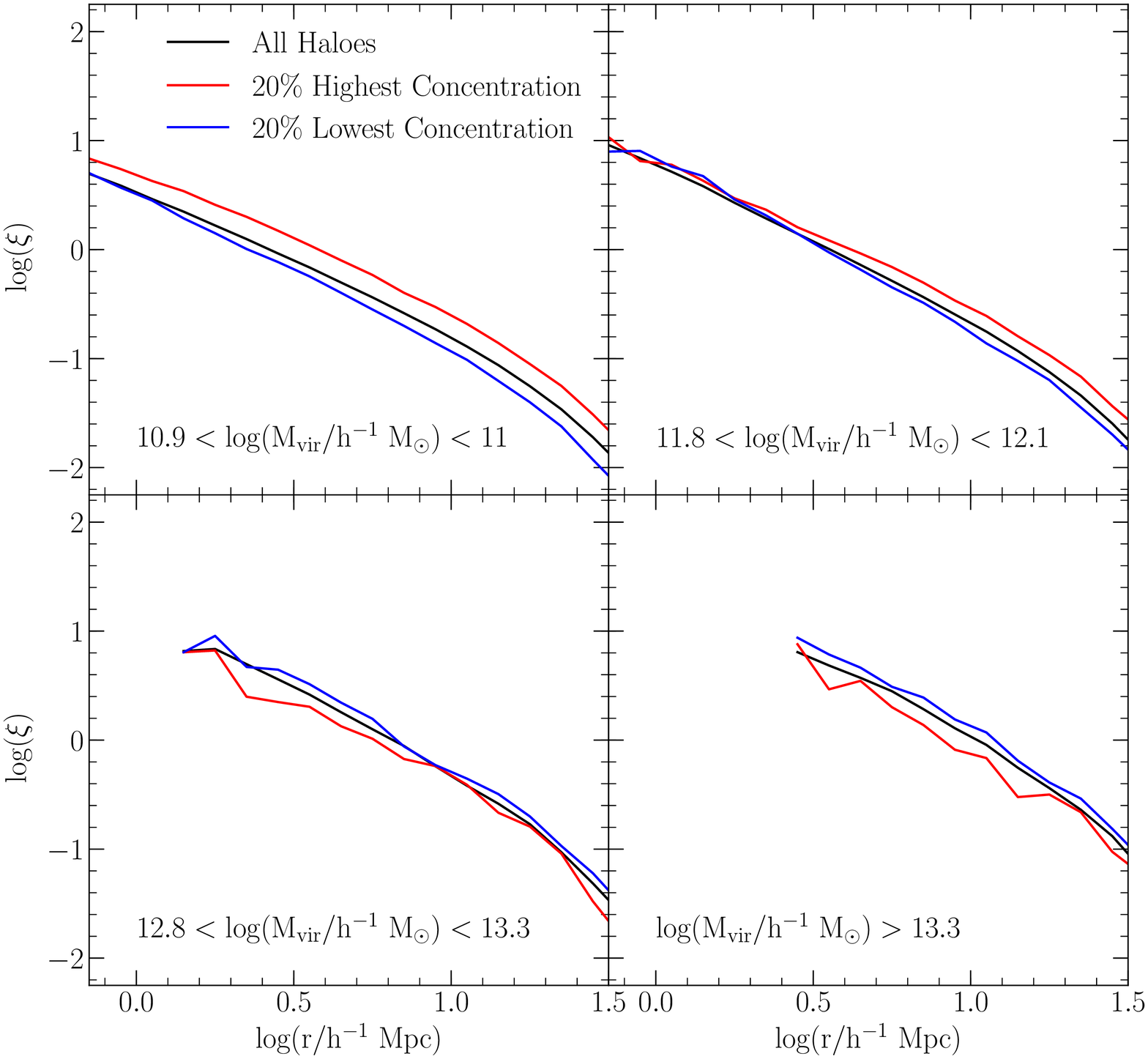}
\hspace{0.4cm}
\includegraphics[width=0.48\textwidth]{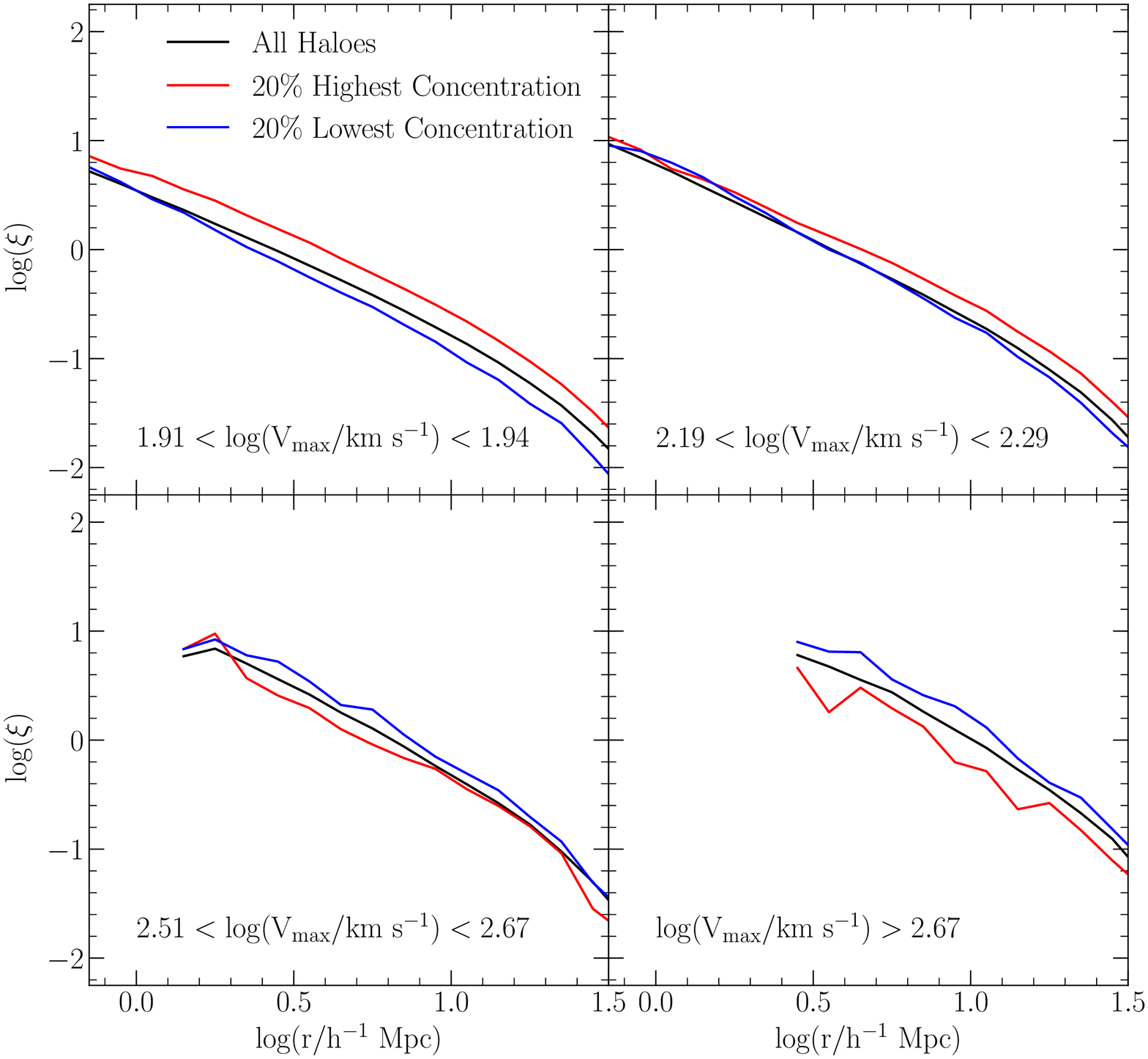}
\caption{The correlation function of halos in the Millennium WMAP7 simulation
and its dependence on concentration, for samples of fixed $\mvir$ (left) and 
fixed $\vmax$ (right).} 
\label{fig:HAB}
\end{figure*}

We examine the concentration dependence of {\it halo} clustering in
the simulation,  namely halo assembly bias. Following \citet{Gao05}, we bin 
the halos in discrete bins of halo mass and calculate the auto-correlation 
function of these halo samples and of the 20\% most concentrated and 20\% 
least concentrated halos.  These are shown in the left-hand side of
Figure~\ref{fig:HAB}, and exhibit the well-studied concentration-dependent
halo clustering (e.g., \citealt{Wechsler06,Gao07,Mao17}). We see that
for relatively-low halo masses, more concentrated halos are more clustered 
than less concentrated halos. This trend is the strongest for our lowest mass
bin, decreases with increasing mass, then reverses sense and continues to
increase in amplitude for the most massive halos.

\begin{figure}
\includegraphics[width=0.48\textwidth]{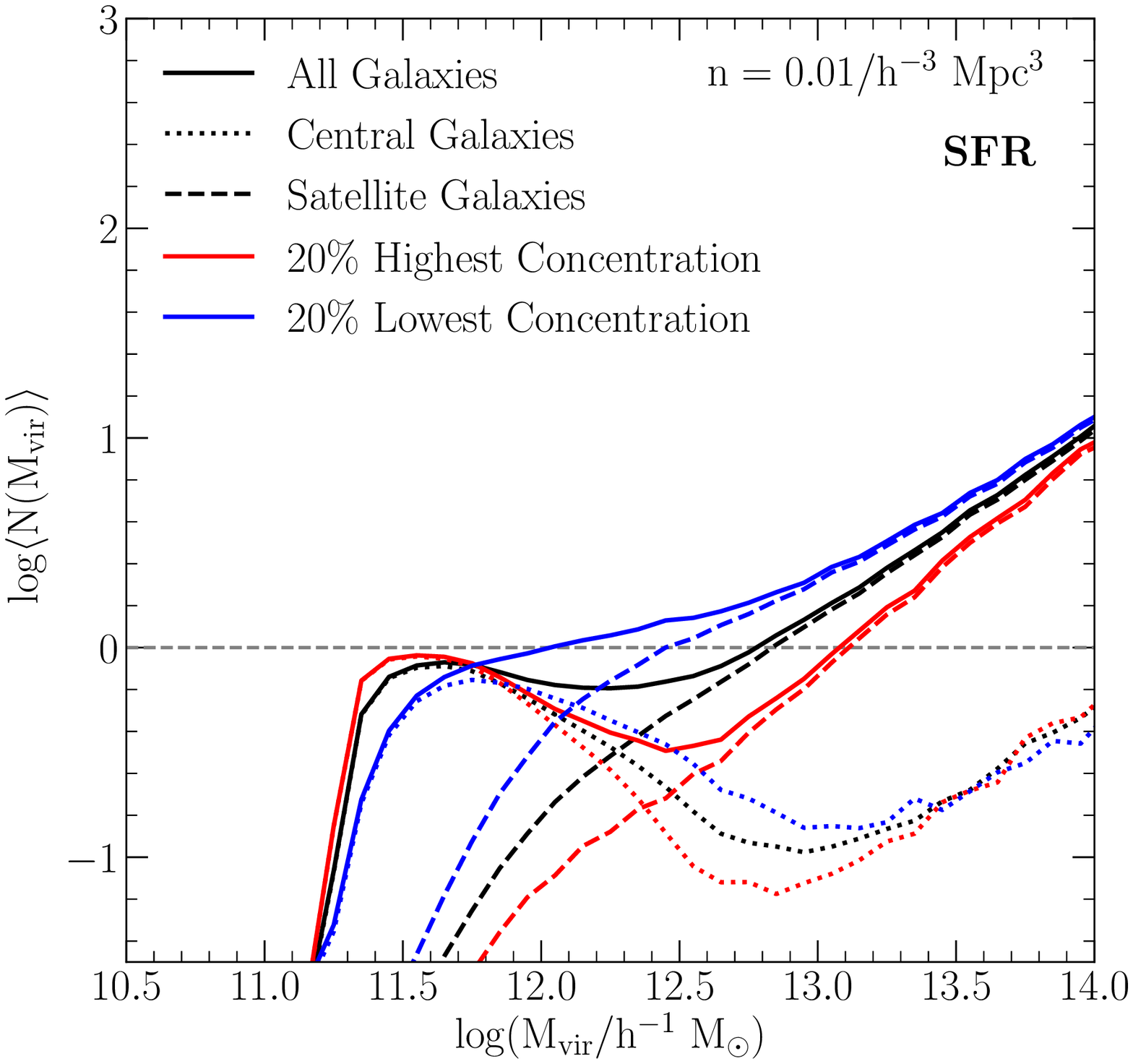}
\hspace{0.4cm}
\includegraphics[width=0.48\textwidth]{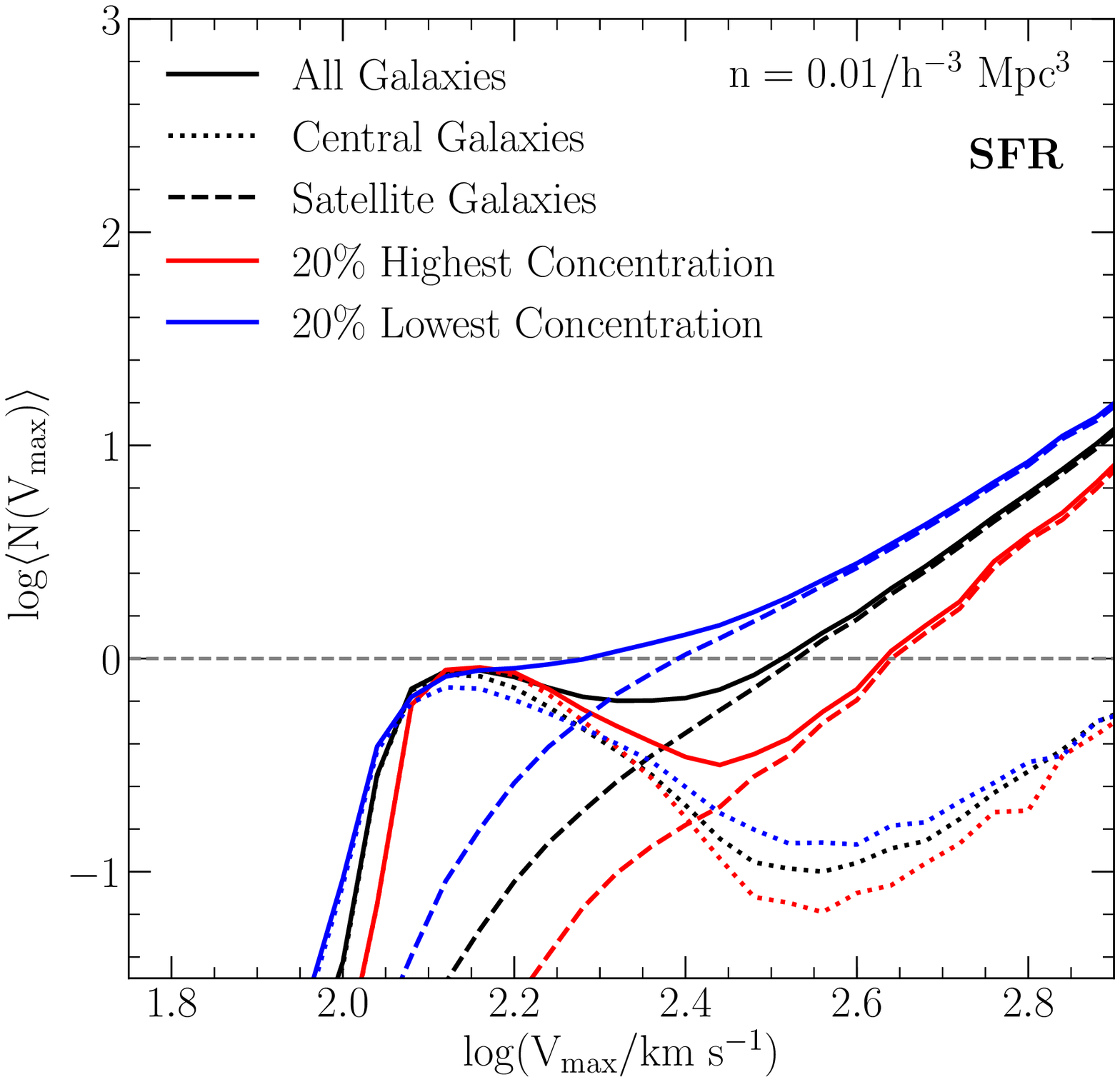}
\caption{Occupancy variation with concentration for a SFR-selected sample
corresponding to a number density of $n=0.01 \hmpcc$, shown for $\mvir$ 
(left) and $\vmax$ (right).}
\label{fig:SFR_HOD}
\end{figure}

\begin{figure}
\centering
\includegraphics[width=0.48\textwidth]{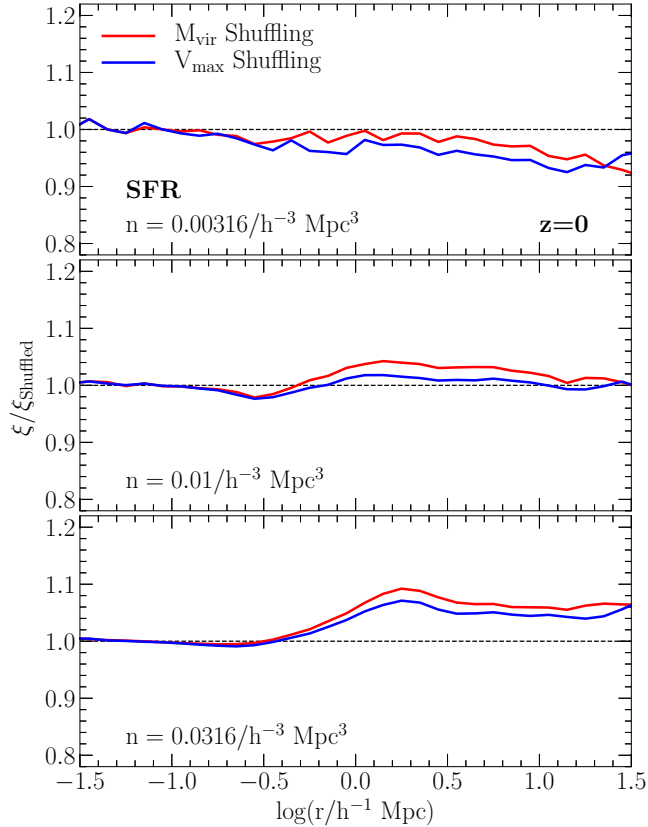}
\caption{Galaxy assembly bias measurements for $\mvir$ (red) and $\vmax$ 
(blue), for the three SFR-selected fixed number density samples.}
\label{fig:SFR_GAB}
\end{figure}

\begin{figure}
\includegraphics[width=0.48\textwidth]{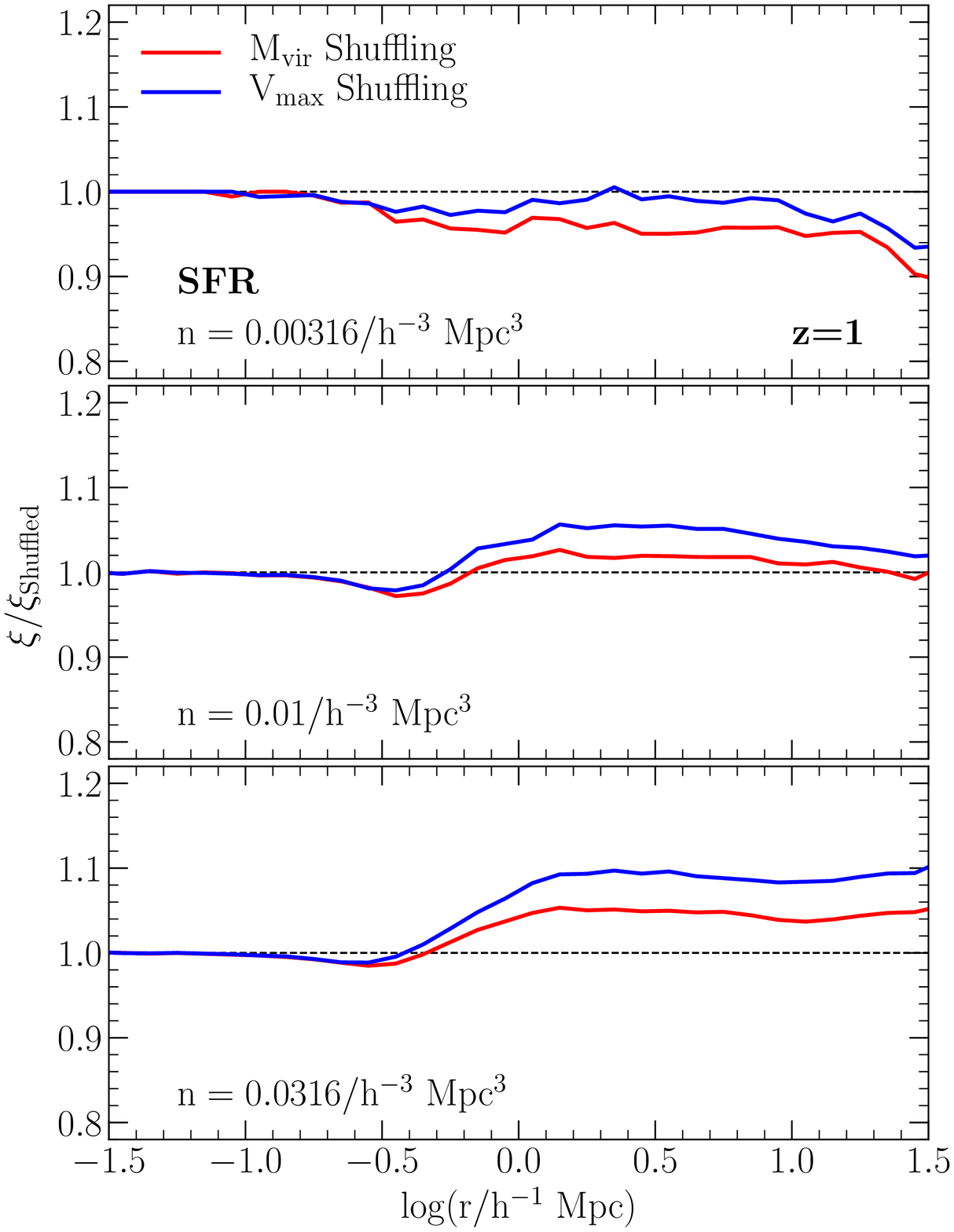}
\hspace{0.4cm}
\includegraphics[width=0.48\textwidth]{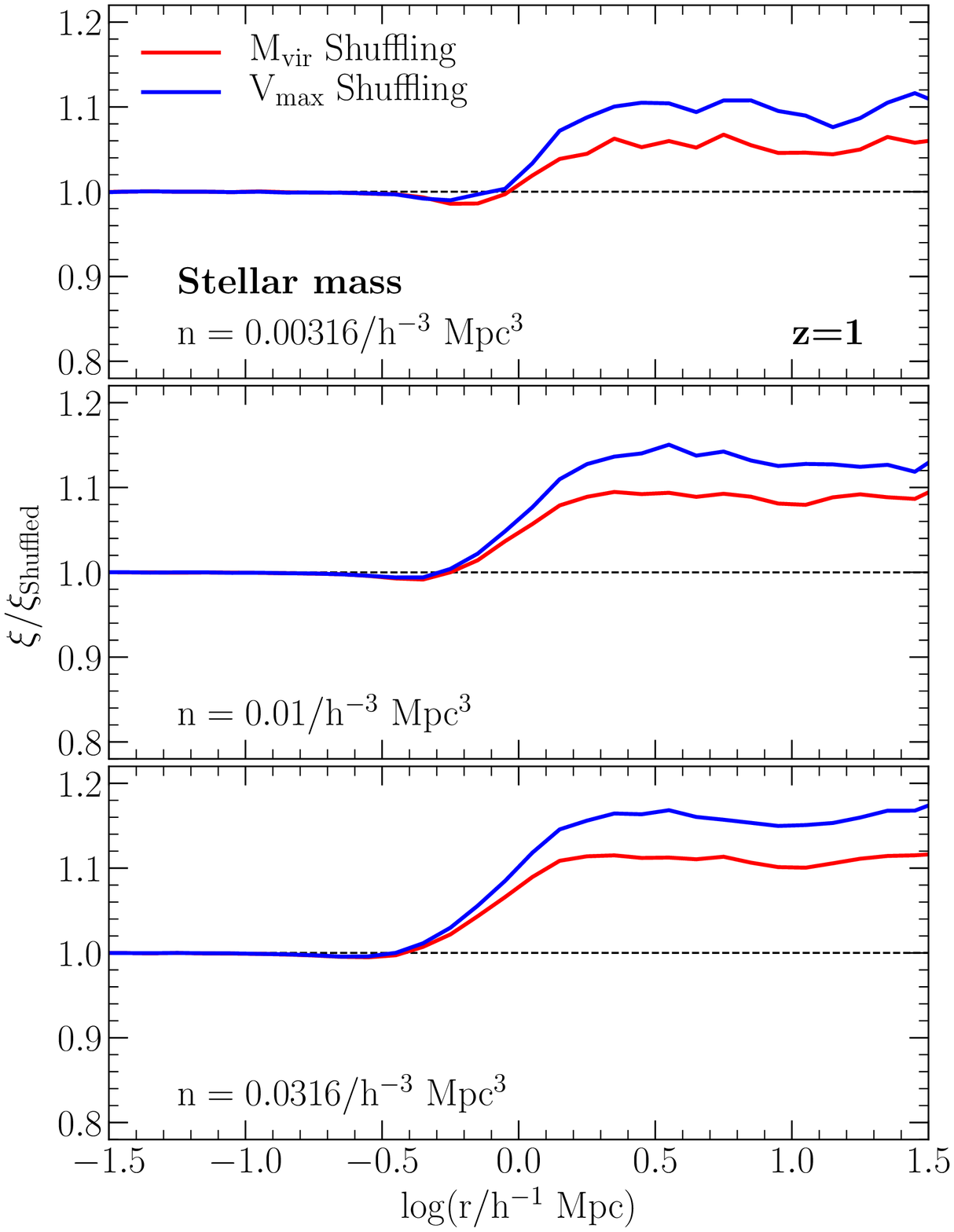}
\caption{Galaxy assembly bias measurements for $\mvir$ (red) and $\vmax$ 
(blue) at $z=1$, shown for the SFR-selected samples (left) and stellar mass
selected samples (right).}
\label{fig:GAB_z1}
\end{figure}

The right-hand side of Figure~\ref{fig:HAB} repeats this analysis, but now in
fixed bins of $\vmax$.  The bins are chosen to roughly match the halo mass bins
according to the relation between $\vmax$ and $\mvir$ (Fig.~\ref{fig:VmaxMvir}) 
and have identical number densities. We find very similar results for the 
concentration-dependent halo clustering as a function of $\vmax$, with 
comparable assembly bias amplitudes in all bins and a reversal of the trend 
at larger $\vmax$ values. Our results are in good agreement with \citet{Sato19}
who find a similar behavior with a reversal of the halo assembly bias effect at 
$\mvir \sim 10^{13} h^{-1} {\rm M}_\odot$ and at $\vmax \sim 330 {\rm km s}^{-1}$,
using the MultiDark suite of simulations \citep{Klypin16}.  
For completeness, we also investigate the concentration-dependence of 
clustering for fixed bins of $\vpeak$, the peak value of $\vmax$ across the
halos' assembly history (not shown here).  We find nearly identical results 
for $\vpeak$ as for $\vmax$.  We conclude that switching from $\mvir$ to 
either $\vmax$ or $\vpeak$ has no significant impact on halo assembly bias.

\vspace{0.1cm}
\section{B. Results for SFR Selected Samples and for Redshift $z=1$}
\label{Sec:SFRz}

We report here an analogous analysis to the one in the main part of the paper
but performed for galaxy samples selected by their SFR, which may be 
relevant for galaxy selections of upcoming surveys. We use the same three
number densities. Figure~\ref{fig:SFR_HOD} shows the occupancy variation
for one representative case of the $n=0.01 \hmpcc$ number density sample.
The left-hand side shows the standard occupation functions as a function of
halo mass. This is the same case presented in Fig.~7 of C19.
Of note is the characteristic shape of the occupation function, which is
different than that of stellar mass selected samples, due to the paucity
of star-forming galaxies residing as centrals in massive halos. The 
occupancy variation for SFR-selected samples is similar to that of stellar 
mass selected ones, with more concentrated halos preferentially hosting
central galaxies at the ``knee'' of the occupation and having fewer satellites.
In the ``dip'' of the centrals occupation at higher mass, a smaller fraction 
of the more concentrated halos tends to host star-forming central galaxies.

The right panel of  Fig.~\ref{fig:SFR_HOD} shows the occupancy variation
in the $\vmax$ case.  In this case, the occupancy variation for the central 
galaxies at the ``turnover'' reverses sense, with low-concentration halos 
starting to host central galaxies at lower halo mass.  The variations in the 
central occupation at higher halo masses remain similar, but perhaps slightly
reduced, while the satellite occupancy variations increase.

The net impact on galaxy clustering is shown in Figure~\ref{fig:SFR_GAB},
where we show again the ratio of correlation function to that of shuffled
samples where the galaxy content was randomly reassigned to halos of the
same $\mvir$ or $\vmax$, which effectively erases the occupancy variations. 
The overall changes with number density are consistent with those found by 
C19. We find that for all SFR-selected samples, this ratio slightly
decreases for $\vmax$ with respect to the one for $\mvir$, likely due to
the preferential occupation of centrals in low-concentration halos. 
Hence we find that, for all cases, the correlation function of the $\vmax$ 
shuffled samples is larger than that of the $\mvir$ shuffled one.  The 
resulting impact on GAB, however, changes with number density (i.e., SFR 
threshold). For the highest number density sample, utilizing $\vmax$ slightly 
decreases GAB (defined as the deviation from a ratio of unity); for the middle 
number density it nearly diminishes the GAB signal; while for the lowest 
number density it increases the GAB effect.  Thus it is hard to draw any 
conclusions on the usefulness of switching to $\vmax$.

Finally, in the left-hand side of Figure~\ref{fig:GAB_z1} we explore the GAB 
signatures at a higher redshift of $z=1$. At $z=1$ we find that the clustering 
ratio is now larger for $\vmax$ than for $\mvir$ for all cases, or rather that 
the clustering of the $\vmax$-shuffled sample is lower than that of the
$\mvir$-shuffled sample. And, once again, the specific impact on GAB depends 
on the number density.
For completeness, we also show in the right-hand side of Fig.~\ref{fig:GAB_z1}
the galaxy assembly bias results for the stellar mass selected samples 
at $z=1$. We see that in this case as well, for all number densities, the
clustering with respect to the $\vmax$-shuffled sample is increased relative 
to the $\mvir$ case, resulting in increased GAB signatures.
Overall, the results presented here strengthen our conclusion that, despite its
claimed ``potential'', $\vmax$ is unable to encapsulate galaxy assembly bias 
effects. 

\begin{figure}
\includegraphics[width=0.48\textwidth]{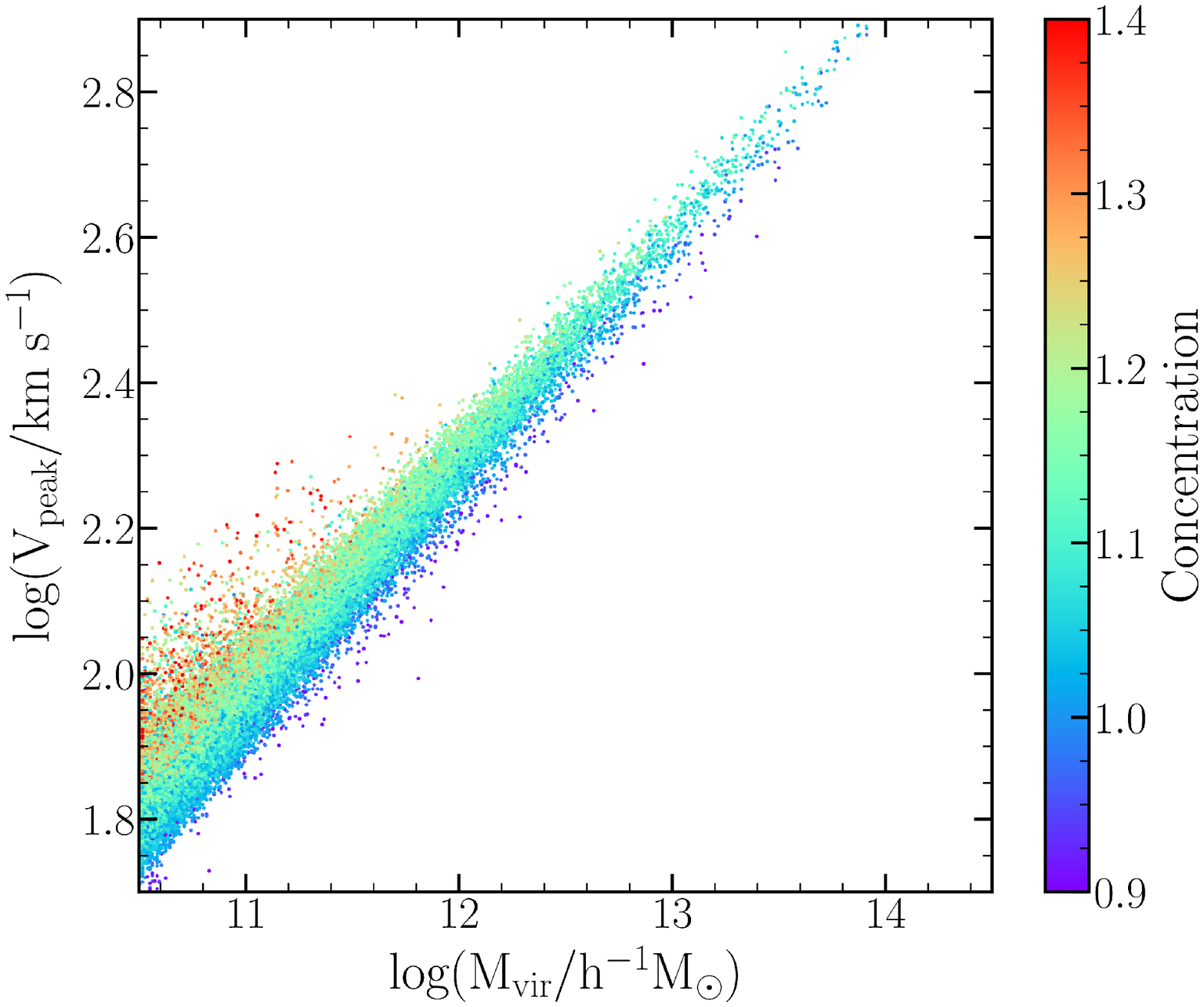}
\hspace{0.4cm}
\includegraphics[width=0.48\textwidth]{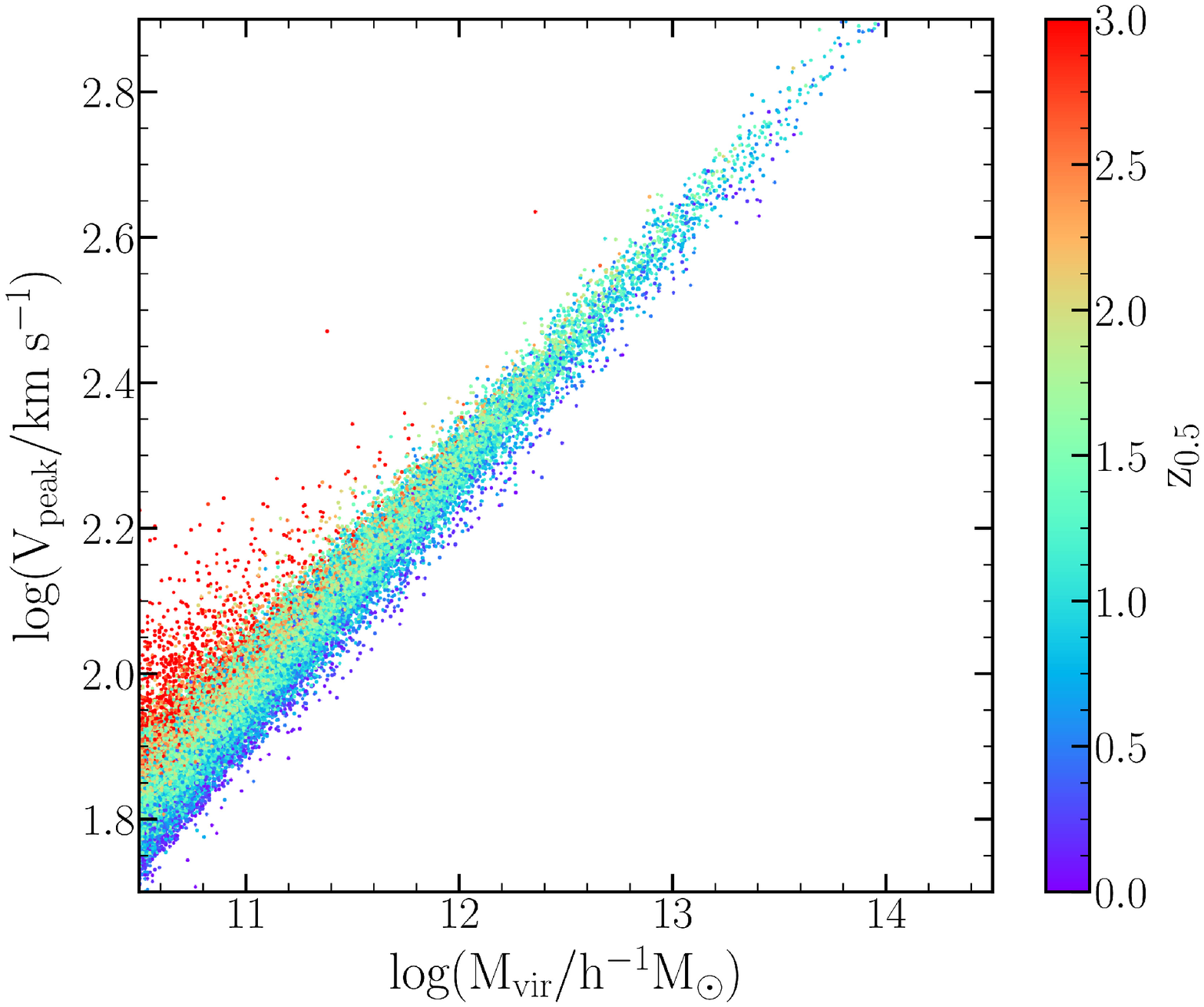}
\caption{The relation between $\vpeak$  and $\mvir$ color coded by halo 
concentration (left) and by halo formation time (right). These are the analog 
of Fig.~\ref{fig:VmaxMvir} but for $\vpeak$ instead of $\vmax$. For clarity,
we plot a representative (randomly chosen and ordered) $1\%$ of the halos.}
\label{fig:VpeakMvir}
\end{figure}

\begin{figure}
\centering
\includegraphics[width=0.48\textwidth]{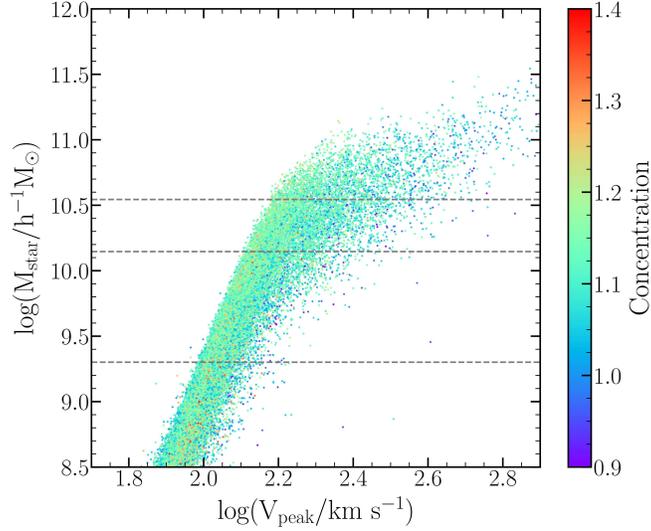}
\caption{The relation between the stellar mass of central galaxies as a 
function of host $\vpeak$, color coded by concentration. The dashed lines 
mark the stellar mass thresholds defining our three samples.  This figure 
complements Fig.~\ref{fig:SMHM}. Once again we only plot $1\%$ of the galaxies.}
\label{fig:SMHM_Vpeak}
\end{figure}

\begin{figure}
\includegraphics[width=0.96\textwidth]{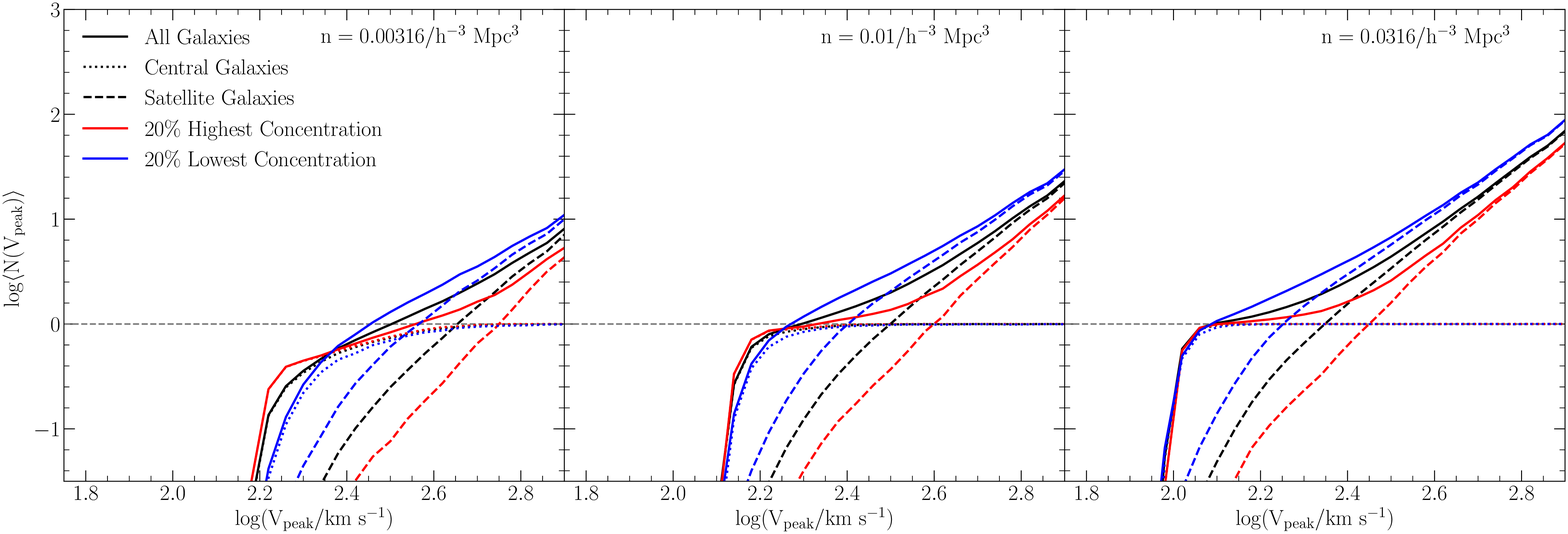}
\caption{The same as Figures~3 and 4 but now for $\vpeak$, namely the halo 
occupation functions for $\vpeak$ showing the variations with concentration, 
for the three number density samples.}
\label{fig:HOD_Vpeak}
\end{figure}

\vspace{0.1cm}
\section{C. Additional results with $\vpeak$}
\label{Sec:Vpeak}

In this appendix we provide supplementary information regarding using $\vpeak$,
the maximum value of $\vmax$ for each halo over cosmic history, as our proxy 
for halo mass. Figure~\ref{fig:VpeakMvir} presents the relation of 
$\vpeak$ to $\mvir$ color coded either by halo concentration (left) or 
formation time (right).  These relations can be compared to the analogous 
ones for $\vmax$ shown in Fig.~\ref{fig:VmaxMvir}.  We find that the relations 
are similar, but with a larger scatter in the $\vpeak$ case, especially for 
the low $\mvir$ / low $\vpeak$ range.  This arises from the scatter between
$\vpeak$ and $\vmax$, primarily for low-mass halos where $\vpeak$ varies 
from $\vmax$ and extends to larger values. It is noteworthy that the low-mass 
halos with the largest $\vpeak$ values tend to have higher concentrations and,
strikingly, earlier formation times.  This supports our hypothesis that these
are splashback halos that had a larger mass in the past.  

Figure~\ref{fig:SMHM_Vpeak} shows the relation between stellar mass and
$\vpeak$ for the SAM central galaxies. This relation is similar to the 
analogous one for $\vmax$ shown in Fig.~\ref{fig:SMHM}. However, it exhibits
less secondary dependences on concentration -- note the lack of noticeable
extremes of the most concentrated halos -- in particular for lower values
of stellar mass.  This contributes to the slight improvement in GAB seen 
in Fig.~\ref{fig:Vpeak} and its dependence on number density.

Finally, Figure~\ref{fig:HOD_Vpeak} shows the occupancy variations with 
concentration for an HOD computed as a function of $\vpeak$. Again, we find
very similar results to the occupancy variations with $\vmax$ 
(Fig.~\ref{fig:HOD} and \ref{fig:HODdiffn}). The centrals occupancy variation 
improves (i.e., decreases) to a varying degree for the different samples while 
the satellites occupancy variation increases with respect to the ``standard'' 
occupancy variations with $\mvir$.  There are slight differences in the level 
of occupancy variation for $\vpeak$ with respect to those for $\vmax$, with
the highest number density sample (lowest stellar mass threshold) exhibiting
now nearly diminished centrals occupancy variation, likely correlated to the
relatively bigger improvement seen for the GAB measurement in that case 
(bottom panel of Fig.~\ref{fig:Vpeak}).

\end{document}